\documentclass[12pt,letterpaper]{article}
\pdfoutput=1

\usepackage{amsmath,amssymb,calc, amsthm,bbm, epsfig,psfrag, mathtools,comment}

\usepackage{dsfont}
\usepackage{mathrsfs}
\usepackage{esvect}
\usepackage{graphicx, enumerate, enumitem}
\usepackage{float}

\usepackage[numbers,sort&compress]{natbib}

\usepackage[dvipsnames]{xcolor}
\usepackage{tikz}
\usepackage{tikz-cd}
\usepackage{tensor}

\usepackage[utf8]{inputenc} 
\definecolor{azure}{rgb}{0.0, 0.5, 1.0}
\definecolor{darkblue}{rgb}{0.15,0.35,0.7}
\definecolor{reddish}{rgb}{0.65, 0.2, 0.2}
\definecolor{brandeisblue}{rgb}{0.0, 0.44, 1.0}
\definecolor{ceruleanblue}{rgb}{0.16, 0.32, 0.75}
\definecolor{indigo(dye)}{rgb}{0.0, 0.25, 0.42}
\usepackage[linktocpage=true]{hyperref}
\hypersetup{
colorlinks=true,
citecolor=ceruleanblue,
linkcolor=ceruleanblue,
urlcolor=ceruleanblue,
pdfauthor={},
pdftitle={},
pdfsubject={}
}

\newcommand{\overbar}[1]{\mkern 1.5mu\overline{\mkern-1.5mu#1\mkern-1.5mu}\mkern 1.5mu}

\newcommand{\ov}[1]{\overbar{#1}}

\newcommand{\cL}{{\cal L}}

\newcommand{\fg}{\mathfrak g}

\newcommand{\nl}{\\[0.4em]}
\newcommand{\ad }{{\rm ad}}

\DeclareMathOperator{\tr}{\text{tr}}

\DeclareFontEncoding{LS1}{}{}
\DeclareFontSubstitution{LS1}{stix}{m}{n}
\DeclareSymbolFont{stixsymbols}{LS1}{stixscr}{m}{n}
\SetSymbolFont{stixsymbols}{bold}{LS1}{stixscr}{b}{n}
\DeclareMathSymbol{\kay}{\mathalpha}{stixsymbols}{"6B}
\DeclareMathSymbol{\hay}{\mathalpha}{stixsymbols}{"68}

\makeatletter
\renewcommand\section{\@startsection {section}{1}{\z@}%
                               {-3.5ex \@plus -1ex \@minus -.2ex}
                               {2.3ex \@plus.2ex}%
                               {\normalfont\large\bfseries}}
\renewcommand\subsection{\@startsection{subsection}{2}{\z@}%
                                 {-3.25ex\@plus -1ex \@minus -.2ex}%
                                 {1.5ex \@plus .2ex}%
                                 {\normalfont\bfseries}}
\makeatother

\let\non\nonumber

\makeatletter
\newcommand*\bigcdot{\mathpalette\bigcdot@{.5}}
\newcommand*\bigcdot@[2]{\mathbin{\vcenter{\hbox{\scalebox{#2}{$\m@th#1\bullet$}}}}}
\makeatother
\newcommand{\deq}{\stackrel{\bigcdot}{=}}
\newcommand{\dapp}{\stackrel{\bigcdot}{\approx}}

\newfont{\goth}{ygoth.tfm scaled 1200}                   

\numberwithin{equation}{section}

\newcommand{\del}{\partial}

\newcommand {\cA}{{\cal A}}
\newcommand {\cB}{{\cal B}}
\newcommand {\cC}{{\cal C}}

\newcommand {\cK}{{\cal K}}

\newcommand {\cO}{{\cal O}}

\newcommand {\cR}{{\cal R}}

\def\a{\alpha}
\def\b{\beta}

\newcommand{\bd}{{\dot{\beta}}}

\newcommand{\bea}{\begin{eqnarray}}
\newcommand{\eea}{\end{eqnarray}}
\newcommand{\ba}{\begin{array}}
\newcommand{\ea}{\end{array}}

\def\double #1{#1{\hbox{\kern-2pt $#1$}}}

\newcommand{\bsubeq}{\begin{subequations}}
\newcommand{\esubeq}{\end{subequations}}

\def\tr{{\rm tr}}

\allowdisplaybreaks

\begin{document}
\begin{titlepage}
\begin{flushright}
\today
\end{flushright}
\vspace{5mm}

\begin{center}
{\Large \bf Relating auxiliary field formulations
\\ 
of $4d$ duality-invariant and $2d$ integrable field theories}
\end{center}

\begin{center}

{\bf
Nicola Baglioni${}^{a}$,
Daniele Bielli${}^{b,c}$,
Michele Galli${}^{a}$,\\
Gabriele Tartaglino-Mazzucchelli${}^{a}$
}
\\
\vspace{5mm}

\footnotesize{
${}^{a}$
{\it 
School of Mathematics and Physics, University of Queensland,
\\
 St Lucia, Brisbane, Queensland 4072, Australia}
}
\\~\\
\footnotesize{
${}^{b}$
{\it 
Asia Pacific Center for Theoretical Physics (APCTP), \\ 
Postech, Pohang 37673, 
Korea}
}
\\~\\
\footnotesize{
${}^{c}$
{\it 
High Energy Physics Research Unit, Faculty of Science \\ 
Chulalongkorn University, Bangkok 10330, Thailand
}
\vspace{2mm}
~\\
\texttt{
n.baglioni@uq.edu.au,
d.bielli4@gmail.com,
m.galli@uq.edu.au,
g.tartaglino-mazzucchelli@uq.edu.au}
}\\
\vspace{2mm}

\end{center}

\begin{abstract}
\baselineskip=14pt
Auxiliary field techniques have recently gained interest in four-dimensional non-linear electrodynamics and two-dimensional integrable sigma models. In these settings, coupling a suitable ``seed'' theory to auxiliary fields provides a powerful mechanism to generate infinite families of models while preserving key dynamical properties, such as electromagnetic duality invariance in four dimensions and classical integrability in two dimensions. Deformations induced through auxiliary fields are closely related to $T\bar{T}$-like deformations and, in two dimensions, also to their higher-spin generalisations. In this paper, we analyse and clarify the relations between different auxiliary field formulations in two and four dimensions, showing how they are governed by Legendre transformations of the interaction functions combined with appropriate field redefinitions. In four-dimensional electrodynamics, we establish a correspondence between the auxiliary field model of Russo and Townsend and the Ivanov--Zupnik formalism. In two dimensions, we develop the analogue of the Ivanov--Zupnik $\mu$-frame to deform Principal Chiral, symmetric-space, non-Abelian T-dual, and (bi-)Yang-Baxter sigma models. We discuss how integrability is preserved and use properties of the $\mu$-frame to further extend known families of integrable deformations.

\end{abstract}
\vspace{5mm}

\vfill
\end{titlepage}

\renewcommand{\thefootnote}{\arabic{footnote}}
\setcounter{footnote}{0}

\tableofcontents{}
\vspace{0.5cm}
\bigskip\hrule

\section{Introduction}\label{sec:intro}

In recent years, there has been a growing interest in various classes of deformations of field theories with the discovery of new links between models in various spacetime dimensions. Our paper uses and extends recent relationships that have emerged between areas such as four-dimensional ($4d$) duality-invariant electrodynamics, (irrelevant) deformations of two-dimensional ($2d$) quantum field theories, and deformations of integrable sigma models. An important ingredient in these developments is the emergence of equivalent and unifying structures governing non-linear Lagrangians, often encoded in remarkably simple functional relations which can be solved by employing suitable choices of (auxiliary field) variables.

A paradigmatic example is provided by non-linear electrodynamics in four dimensions. It has been shown that continuous, electromagnetic (non-linear) duality-invariance severely constrains the form of admissible interactions \cite{Gaillard:1981rj} -- see also
\cite{BB:1983,Gibbons:1995cv,Gaillard:1997rt,Gaillard:1997zr,Hatsuda:1999ys,Kuzenko:2000uh,Ivanov:2002ab,Ivanov:2003uj} for an incomplete list of early papers on this subject as well as the more recent pedagogical review \cite{Sorokin:2021tge}.
A well-known example of a one-parameter family of duality-invariant models of non-linear electrodynamics is the Maxwell-Born-Infeld theory, which has played an important role in string theory for decades -- see, e.\,g., \cite{Tseytlin:1999dj}.
Recently, duality-invariant models of non-linear electrodynamics have attracted further attention following the surprising discovery of the ModMax theory \cite{Bandos:2020jsw}, a one-parameter family of (classically) conformal and duality-invariant deformation of Maxwell's theory which interpolates between free electrodynamics and strongly non-linear and non-analytic regimes. In parallel, significant progress has been made in the study of irrelevant deformations of two-dimensional quantum field theories, most notably the $T\bar T$ deformation and its generalisations -- see \cite{Zamolodchikov:2004ce,Cavaglia:2016oda,Smirnov:2016lqw} for the seminal papers, as well as \cite{Jiang:2019epa,Ferko:2021loo,He:2025ppz} for references on their developments and pedagogical reviews. These deformations exhibit a number of remarkable properties, including the solvability of several observables and close ties to integrable systems.

From a Lagrangian perspective, it was shown that the Born-Infeld Lagrangian satisfies a $4d$ analogue of a $T\bar{T}$ deformation \cite{Conti:2018jho}, while it was more recently shown that ModMax satisfies a so-called root-$T\bar T$ deformation driven by a non-analytic though classically marginal operator constructed out of the square root of a squared combination of the energy momentum tensor \cite{Babaei-Aghbolagh:2022uij,Ferko:2022iru,Conti:2022egv,Ferko:2022cix, Ferko:2023wyi} -- see also \cite{Ferko:2022cix,Borsato:2022tmu,Conti:2022egv,Babaei-Aghbolagh:2022leo,Ferko:2024zth} for extensions to $2d$ and $6d$. The work \cite{Ferko:2023wyi} has, in particular, proven the remarkable fact that generic families of duality-invariant models of $4d$ electrodynamics necessarily satisfy a $T\bar{T}$-like flow. Moreover, in \cite{Ferko:2023wyi} auxiliary field approaches based on either an auxiliary real two-form field or a complex scalar field were employed to efficiently engineer such generic $T\bar{T}$-like deformations in four dimensions. Similar techniques based on a Lie algebra valued $2d$ vector auxiliary field were then engineered for generic $T\bar{T}$-like, and even higher-spin (like the ones of Smirnov and Zamolodchikov of \cite{Smirnov:2016lqw}), deformations of two-dimensional integrable sigma models \cite{Ferko:2024ali,Bielli:2024khq,Bielli:2024ach, Bielli:2024fnp, Bielli:2024oif, Cesaro:2024ipq,Fukushima:2024nxm,Bielli:2025uiv,Ferko:2025bhv}. This viewpoint has proven particularly useful in clarifying the relation between stress-tensor deformations, (classical) $2d$ integrability, and $d>2$ duality invariant models.

It is worth noting that other methods for engineering $T\overbar{T}$ in arbitrary dimensions make use of an auxiliary metric (vielbein) \cite{Coleman:2019dvf,Tolley:2019nmm,Conti:2022egv,Morone:2024ffm,Babaei-Aghbolagh:2024hti,Tsolakidis:2024wut,Ferko:2024yhc,Brizio:2024arr}. This method, which is often referred to as gravitational dressing, connects the study of deformations of quantum field theories to $2d$ (topological) gravity \cite{Dubovsky:2017cnj,Cardy:2018sdv,Dubovsky:2018bmo,Conti:2018tca,Conti:2019dxg,Caputa:2020lpa,Conti:2022egv,Bhattacharyya:2023gvg}, and the broad literature on $d$-dimensional models of massive gravity \cite{Tolley:2019nmm,Ran:2024vgl,Morone:2024ffm,Babaei-Aghbolagh:2024hti,Tsolakidis:2024wut,Ferko:2024yhc,Brizio:2024arr,Morone:2024sdg,Callebaut:2025thw}. An advantage of this approach is its potential universality as, in principle, it applies to any field theory that can be minimally coupled to gravity. Moreover, integrable structures are well understood in the case of $T\bar{T}$ (however, not for more general $T\bar{T}$-like cases), where it is known that the gravitational dressing takes the form of a field-dependent change of coordinates. For more general $T\bar{T}$-deformations, including higher-spin ones, the vector auxiliary field approach of \cite{Ferko:2024ali,Bielli:2024ach,Bielli:2025uiv} has shown remarkable simplicity in the description of classical integrability.

Auxiliary fields are, of course, widely used in quantum field theory, including in off-shell supersymmetry \cite{Gates:1983nr,Wess:1992cp,Buchbinder:1998qv}, chiral fields in various dimensions \cite{Siegel:1983es,Pasti:1995tn,Pasti:1996vs,Pasti:1997gx,Ivanov:2014nya,Buratti:2019cbm,Buratti:2019guq,Sen:2015nph,Sen:2019qit,Townsend:2019koy,Mkrtchyan:2019opf,Bansal:2021bis,Avetisyan:2021heg,Avetisyan:2022zza,Evnin:2022kqn,Arvanitakis:2022bnr,Ferko:2024zth,Sorokin:2025ezs,Kuzenko:2025jgk,Cederwall:2025ywy,Hutomo:2025dfx}, and, as already mentioned, theories of $4d$ duality-invariant non-linear electrodynamics\footnote{Auxiliary field techniques have also been developed for supersymmetric duality-invariant non-linear electrodynamics \cite{Kuzenko:2013gr,Ivanov:2013ppa} as well as (supersymmetric) gauge theories with arbitrary spin fields \cite{Kuzenko:2021qcx,Kuzenko:2021pqm}.} \cite{Hatsuda:1999ys,Ivanov:2002ab,Ivanov:2003uj,Ivanov:2013jba,Ferko:2023wyi,Kuzenko:2024zra,Russo:2024xnh,Russo:2024llm,Russo:2025fuc,Babaei-Aghbolagh:2025uoz,Kuzenko:2025gvn, Chen:2025ndc} as well as integrable $T\overbar{T}$-like deformations of integrable sigma models \cite{Ferko:2024ali, Bielli:2024khq, Bielli:2024ach, Bielli:2024fnp, Bielli:2024oif, Bielli:2025uiv, Ferko:2025bhv, Cesaro:2024ipq, Fukushima:2024nxm, Babaei-Aghbolagh:2025uoz, Babaei-Aghbolagh:2025hlm,Fukushima:2025tlj,Fukushima:2026gan}.

The striking connection between these research directions is the fact that the same Courant--Hilbert (CH) functional equation governing duality-invariant electrodynamics also controls the integrability of a broad class of two-dimensional sigma models. In the context of Principal Chiral Models (PCMs), it has been shown that Lagrangians depending on specific combinations of currents satisfy integrability conditions precisely when they solve a CH-type equation \cite{Borsato:2022tmu,Ferko:2023wyi}. This observation provided a unifying framework in which duality invariance in four dimensions and integrability in two dimensions arise from closely related mathematical structures and remains a source of inspiration to explore new corners of this correspondence.

The purpose of this work is to further develop and clarify the various relationships among duality-invariant $4d$ electrodynamics, $T\bar{T}$-like deformations, and integrable sigma models. On the one hand, we aim to relate the recent auxiliary field solution of the Courant-Hilbert equation, that was developed in the work \cite{Russo:2025fuc}, to the Ivanov-Zupnik auxiliary field formulation of general duality-invariant theories of electrodynamics \cite{Ivanov:2002ab,Ivanov:2003uj}. On the other hand, a second aim of our paper is to extend the recent auxiliary field deformations of integrable sigma models \cite{Ferko:2024ali, Bielli:2024khq, Bielli:2024ach, Bielli:2024fnp, Bielli:2024oif, Bielli:2025uiv, Ferko:2025bhv}.  Central to our analysis is the use of Legendre transformations to relate different ``frames'', in which the same physical theory admits seemingly different auxiliary field descriptions. In particular, we demonstrate how the $\nu$-frame and $\mu$-frame formulations, originally introduced by Ivanov and Zupnik, naturally bridge various formulations of four-dimensional duality-invariant electrodynamics and two-dimensional integrable deformations.

A key outcome of this perspective is that some integrability properties simplify in suitably chosen frames, where Lie-algebra-valued vector auxiliary fields are traded for scalar ones. This simplification allows us to analyse integrable structures, Lax connections, and Poisson brackets in a systematic way, and to extend previous results to a broader class of deformations involving multiple auxiliary variables.

To provide more detailed context for the motivation and results of our work, we now focus on a particularly useful auxiliary field realisation of these ideas, which has been employed in both four-dimensional electrodynamics and two-dimensional integrable models \cite{Russo:2025fuc, Ivanov:2001ec,Ivanov:2002ab, Ivanov:2003uj, Fukushima:2024nxm}.
Russo and Townsend (RT) have recently put forward a model \cite{Russo:2025fuc} characterised by the presence of a single real scalar auxiliary field $y$, which turns out to be particularly useful in studying various properties of duality-invariant theories of electrodynamics.\footnote{We refer the reader to \cite{Aschieri:2008ns,Kuzenko:2000uh,Sorokin:2021tge} as well as the recent papers \cite{Ferko:2023ruw,Russo:2025fuc} for pedagogical definitions of nonlinear duality-invariance in $4d$ electrodynamics.} The details of the model are encoded into a generic interaction function $\Omega(y)$, which is only a function of the auxiliary field variable $y$ and parametrises families of duality-invariant theories. Using this description, the authors have been able to study (among other properties) 
the causality of these models \cite{Russo:2024llm}, as well as the requirements ensuring the existence of a weak-field analytic expansion \cite{Russo:2025fuc}, in terms of conditions on $\Omega(y)$. The deeper origin of this approach lies in the Courant-Hilbert (CH) solution \cite{CH:1962} to the partial differential equation \cite{Gibbons:1995cv}
\begin{equation}\label{PDE}
\frac{\partial \mathcal L}{\partial U}\frac{\partial \mathcal L}{\partial V}=-1  \qquad \text{with} \qquad \mathcal{L}=\mathcal{L}(U,V) \,\, ,
\end{equation}
known, since the work of Bialynicki-Birula \cite{BB:1983}, to encode the requirement for having an electromagnetic duality-invariant theory. In theories of electrodynamics, the variables $U$ and $V$ are given by
\begin{equation}\label{U-V-def}
U=\frac{1}{2}\left(\sqrt{S^2+P^2}-S\right) \,\, ,
\qquad \qquad 
V=\frac{1}{2}\left(\sqrt{S^2+P^2}+S\right)\,\, .
\end{equation}
Namely, they are functions of the scalar and pseudo-scalar contractions, $S$ and $P$, of the field strength tensor $F_{\mu\nu}$, which in the conventions of \cite{Bandos:2020jsw} can be written as 
\begin{equation}\label{S-P-def}
S=-\frac{1}{4}F^{\mu\nu}F_{\mu\nu}, \qquad  \qquad 
P=-\frac{1}{8}\epsilon^{\mu\nu\rho\sigma}F_{\mu\nu}F_{\rho\sigma} \,\, .
\end{equation}
The CH solution \cite{CH:1962} to the equation \eqref{PDE} can be written as:
\begin{equation}\label{PDE_intro}
\mathcal L= \ell(\tau)-\frac{2 U}{\dot\ell(\tau)} \,\, ,
\qquad \text{with} \qquad 
\tau=V+\frac{U}{\dot\ell(\tau)^2} \,\, ,
\end{equation}
and is implicitly encoded in the single real function $\ell(\tau)$ of the real variable $\tau$. Given a choice of $\ell(\tau)$, the second equation in \eqref{PDE_intro} fixes the auxiliary variable $\tau$ as a function of the physical Lorentz scalars $U$ and $V$, $\tau=\tau(U,V)$. This, in turn, fully characterises the Lagrangian, which is ensured to satisfy \eqref{PDE} and hence the resulting model is invariant under nonlinear SO$(2)$ electromagnetic duality transformations.

It was already noted in \cite{Borsato:2022tmu,Ferko:2023wyi} (see also \cite{Babaei-Aghbolagh:2025uoz}) that the condition \eqref{PDE}, equivalently expressed as the following partial differential equation\footnote{In the $2d$ case, as we will discuss in the main body of the paper, the variable $S$ and $P$ are defined, for example, in terms of the Maurer-Cartan one-form $j$ of the Principal Chiral Model as $S = - \frac{1}{2} \tr \left( j_+ j_- \right)$ and $P^2 = \frac{1}{4} \left( \tr \left( j_+ j_+ \right) \tr \left( j_- j_- \right) - \left( \tr \left( j_+ j_- \right) \right)^2 \right)$.}
\begin{align}
    \mathcal{L}_S^2 - \frac{2 S}{P} \mathcal{L}_S \mathcal{L}_P - \mathcal{L}_P^2 = 1 \, , 
\end{align}
and its solution \eqref{PDE_intro}, naturally arises also in the context of two-dimensional integrable sigma models as an integrability condition ensuring the equivalence of the equations of motion and flatness of a Lax connection. This apparent coincidence has already been exploited, via the formalism of Ivanov and Zupnik (IZ) for four-dimensional electrodynamics \cite{Ivanov:2001ec,Ivanov:2002ab,Ivanov:2003uj}, to construct, as already mentioned, deformations of two-dimensional integrable sigma models first by Ferko and Smith in \cite{Ferko:2024ali} and successively in various other papers \cite{Bielli:2024khq, Bielli:2024ach, Bielli:2024fnp, Bielli:2024oif, Bielli:2025uiv, Ferko:2025bhv}. The IZ approach allows to describe families of electrodynamics and integrable theories by coupling the physical fields to non-scalar auxiliary fields, characterised by an arbitrary interaction function $E$ which does not depend on the physical fields. Although at first glance this perspective might seem quite different from the CH solution, which relies on a single scalar auxiliary field, IZ models can be restricted to depend on specific scalar combinations of the auxiliary fields describing duality-invariant theories in $4d$ and integrable theories in $2d$, as respectively discussed at length in \cite{Ferko:2023wyi} and \cite{Bielli:2024ach}. Starting from this observation one might expect both the Courant-Hilbert solution and the related Russo-Townsend model \cite{Russo:2025fuc} to be able to describe families of integrable theories. This motivates us to explore such relation, which will be made more precise in the rest of this work.

In section \ref{s:4d} we recall the IZ model in $4d$ \cite{Ivanov:2002ab,Ivanov:2003uj} and show how this is connected to the RT model \cite{Russo:2025fuc}. The main observation is that the IZ model, normally expressed in the so-called $\nu$-frame, also enjoys a possibly less known equivalent description, the $\mu$-frame, which can be defined by Legendre transforming the interaction function $E$ when one imposes that it depends on a specific scalar combination of the auxiliary fields \cite{Ivanov:2002ab,Ivanov:2003uj}. Such a combination is exactly the one ensuring duality-invariance of the underlying electrodynamics theory and allows to precisely connect the IZ model in the $\mu$-frame to the RT model.

In section \ref{sec:mu-frame-2d}, we shift our attention from four to two dimensions. The IZ formalism has so far been exploited to construct integrable deformations of two-dimensional sigma models in the $\nu$-frame \cite{Ferko:2024ali, Bielli:2024khq, Bielli:2024ach, Bielli:2024fnp, Bielli:2024oif, Bielli:2025uiv, Ferko:2025bhv} and, analogously to the four-dimensional case, we show that changing perspective to the $\mu$-frame allows to uncover a precise relation to the auxiliary scalar field model recently studied in \cite{Babaei-Aghbolagh:2025hlm, Babaei-Aghbolagh:2025uoz}, which represents the two-dimensional equivalent of the RT model. Along with such relation, we also discuss an extension of the formalism which, starting from the $\nu$-frame in the presence of two independent types of auxiliary field combinations, allows to construct a $\mu$-frame written in terms of two auxiliary real scalar fields.

In section \ref{s:int} we study the integrable structure of $2d$ IZ-like theories in the $\mu$-frame. While the Lax-integrability of the CH solution has already been discussed \cite{Babaei-Aghbolagh:2025hlm}, a complete analysis of its underlying Poisson structure is still missing. Exploiting the correspondence with the $\nu$-frame AFSM \cite{Ferko:2024ali}, we show how the Poisson brackets of the Lax connection in the $\mu$-frame exhibit the non-ultralocal Maillet structure \cite{MAILLET198654} which ensures commutativity of the conserved charges. 

Section \ref{s:ext} deals with the task of extending the $\mu$-frame description from the simplest case of PCMs \cite{Ferko:2024ali} to other classes of $2d$ sigma models to which the IZ-like auxiliary field deformation has been applied in the $\nu$-frame. This specifically includes the non-Abelian T-dual of the PCM \cite{Bielli:2024khq,Bielli:2024ach}, (bi-)Yang-Baxter deformations \cite{Bielli:2024fnp} and symmetric-space sigma models \cite{Bielli:2024oif,Cesaro:2024ipq,Cesaro:2025msv}. This shows how all these models also exhibit a description in terms of a scalar auxiliary field model and, analogously, of a CH solution --- see also \cite{Fukushima:2026gan} for recent developments.

Some computational details are finally collected in a few appendices. In appendix \ref{app_a} we show the equivalence of the equations of motion (EOM) for the auxiliary field PCM in the IZ formalism from the $\nu$ and $\mu$ frame perspectives. Appendix \ref{app_b} provides detailed expressions for the stress-tensors in both frames and appendix \ref{app_C} discusses the detailed equivalence for sigma-models beyond the PCM.

\section{Non-linear electrodynamics (NLED) and auxiliary fields}\label{s:4d}

It has been highlighted in the introduction how the inclusion of auxiliary fields, namely non-dynamical degrees of freedom which can be fixed through their own equations of motion, has been particularly relevant in the context of NLED. As we will discuss below, it has not only allowed for the direct construction of theories generalising Maxwell's electrodynamics, it has also been shown to naturally arise in the context of duality-invariant NLED theories, underlining the existence of a tight connection between duality-invariance and the presence of auxiliary fields. Our main objective throughout this section will be that of exploring such a relation, showing how the auxiliary field formulation arising in the Russo-Townsend formalism from duality-invariance, are related to a certain extension of Maxwell's theory introduced by Ivanov and Zupnik. We will also comment on the relationships between these approaches and the Courant-Hilbert one.

\subsection{The Ivanov-Zupnik (IZ) extension of Maxwell's electrodynamics}\label{subsec:IZ-formalism-4d}
This extension of Maxwell's theory was constructed more than two decades ago by Ivanov and Zupnik (IZ) \cite{Ivanov:2001ec,Ivanov:2002ab,Ivanov:2003uj} by introducing auxiliary fields coupling to the physical field strength tensor $F_{\mu\nu}:=2\partial_{[\,\mu}A_{\nu\,]}$\footnote{Square brackets denote antisymmetrisation of the enclosed indices with the appropriate numerical counting factor: $T_{[\mu_{1}\mu_{2}...\mu_{n}]}=\tfrac{1}{n!}(T_{\mu_{1}\mu_{2}...\mu_{n}}-T_{\mu_{2}\mu_{1}...\mu_{n}}+...)$. Moreover, we refer the reader to \cite{Ferko:2023wyi} for the $4d$ notations used in this section.} and characterised by the presence of an unspecified interaction function which does not depend on the physical degrees of freedom. As shown in the original papers, the formalism naturally exhibits two different formulations, or frames, named after the specific set of auxiliary variables used to describe them.

\paragraph{$\nu$-frame.} In this formulation, the IZ-Lagrangian takes the form
\begin{equation}\label{L_IZ_4d}
\mathcal{L}=\frac{1}{2}(\varphi+\bar\varphi)-2(V\cdot F+\bar V\cdot \bar F)+\nu+\bar\nu+E(\nu,\bar \nu)\, ,
\end{equation}
with the field strength $F_{\mu\nu}$ expressed in spinorial components 
\begin{equation}
F_{\alpha}{}^{\beta}=-\frac{1}{4}(\sigma^\mu)_{\alpha\dot\beta}\,(\tilde \sigma^\nu)^{\dot\beta \beta}F_{\mu\nu}
\qquad \text{and} \qquad
\bar{F}^{\dot\alpha}{}_{\dot\beta}=\frac{1}{4}\,(\tilde \sigma^\mu)^{\dot\a \beta}\,(\sigma^\nu)_{\beta\dot\b} F_{\mu\nu} \,\, ,
\end{equation}
and also appearing in the first term via
\begin{equation}\label{phi-barphi-def}
\varphi =F_{\alpha\beta}F^{\alpha\beta}
\qquad \text{and} \qquad 
\bar\varphi=\bar F_{\dot\alpha\dot\beta}\bar F^{\dot\alpha\dot\beta} \,\,,
\end{equation}
following the notation of \cite{Ferko:2023wyi}. The $\nu$-variables are defined in terms of an antisymmetric auxiliary field $V_{\mu\nu}$, also written in spinorial components
\begin{equation}
\begin{aligned}
    \nu=&~V_{\alpha \beta}V^{\alpha \beta} , \qquad \qquad  V_{\alpha}{}^{\beta}=-\frac{1}{4}(\sigma^\mu)_{\alpha\dot\beta}\,(\tilde \sigma^\nu)^{\dot\beta \beta}V_{\mu\nu}\,,
    \\
    \bar\nu=&~\bar V_{\dot\alpha \dot\beta}\bar V^{\dot\alpha \dot\beta} , \qquad \qquad \bar V^{\dot\alpha}{}_{\dot\beta}=\frac{1}{4}\,(\tilde \sigma^\mu)^{\dot\a \beta}\,(\sigma^\nu)_{\beta\dot\b}\bar V_{\mu\nu}\, ,
\end{aligned}
\end{equation}
and terms mixing physical and auxiliary fields are similarly defined as
\begin{equation}
V\cdot F=V_{\alpha \beta}F^{\alpha \beta} \, ,
\qquad \qquad \bar V\cdot \bar F=\bar V_{\dot\alpha \dot\beta}\bar F^{\dot\alpha \dot\beta}\,.
\end{equation}
Importantly, the auxiliary fields $V_{\a\b}$ and $\bar{V}_{\dot\a\bd}$ only appear in the interaction function $E$ through the specific combinations $\nu$ and $\bar\nu$. Being non-dynamical degrees of freedom, their equations of motion are algebraic equations that can be used to re-express such fields in terms of the physical composite fields $\varphi,\bar\varphi$, and it is immediate to see that when $E=0$ Maxwell electrodynamics is recovered with this procedure.

Throughout our analysis we will use the symbol ``$\deq$'' to indicate equalities holding when the auxiliary field equations of motion are assumed to be satisfied. Similarly, we will use ``$\approx$'' when the physical fields are on shell, and the combined notation ``$\dapp$'' when the equations of motion of both physical and auxiliary fields are used.
For general choices of the function $E$, the equations of motion for $V_{\alpha\beta}$ and $\bar V_{\dot\alpha \dot \beta}$ are:
\begin{equation}\label{F_eom}
F_{\alpha\beta}\deq V_{\alpha\beta} \left(1+E_{\nu}\right)
\qquad \text{and} \qquad 
F_{\dot\alpha \dot \beta}\deq \bar V_{\dot\alpha \dot \beta}\left(1+E_{\bar\nu}\right) \,\, ,
\end{equation}
which imply the Lorentz invariant conditions
\begin{equation}\label{phi_eom}
\varphi\deq \nu \left(1+E_{\nu}\right)^2
\qquad \text{and} \qquad 
\bar\varphi\deq \bar\nu \left(1+E_{\bar\nu}\right)^2 \,\, ,
\end{equation}
where subscripts are used to denote derivatives $f_x=\frac{\partial f}{\partial x}$.
This fact had already been used in \cite{Ivanov:2002ab} to define the so-called $\mu$-frame alternative formulation.

\paragraph{$\mu$-frame.} In this picture the IZ-Lagrangian takes the form
\begin{equation}\label{L_mu_barmu}
    \mathcal{L}=\frac{1}{2}\left[\varphi\frac{(\mu-1)}{(\mu+1)}+\bar\varphi\frac{(\bar\mu-1)}{(\bar\mu+1)}\right]+H(\mu,\bar\mu)\,,
\end{equation}
where ($\mu$, $\bar\mu$) and $H$ are defined from ($\nu$, $\bar\nu$) and $E$ via a (complex) Legendre transformation
\begin{equation}\label{general_legendre}
    H=E-\nu E_\nu-\bar\nu E_{\bar\nu}\,,\quad\qquad \text{with} \quad\qquad \mu=E_{\nu}\,,\qquad \bar\mu=E_{\bar\nu }\, .
\end{equation}
It was already observed in \cite{Ivanov:2002ab} that requiring the theory to be invariant under SO$(2)$ electromagnetic duality rotations amounts to having $E$ only depend on products $\nu \bar \nu$ and similarly $H$ only depend on $\mu\bar\mu$. 

An important difference between the above two perspectives is that while in the $\nu$-frame the auxiliary fields are tensors of the same type as the physical field strength, when moving to the $\mu$-frame they automatically turn into much simpler scalar fields. This observation will play an important role in the rest of this section, where our main goal will be to spell out the precise correspondence between the $\mu$-frame setup and the auxiliary field model recently introduced by Russo and Townsend in  \cite{Russo:2024llm}.

\subsection{The Russo-Townsend (RT) formulation of duality-invariant NLED}\label{subsec:RT-formalism-4d}
In \cite{Russo:2024llm}, the authors studied how duality-invariant theories of electrodynamics can be characterised in terms of the Courant-Hilbert function $\ell(\tau)$, which parametrises the general solution \eqref{PDE_intro} to the differential equation \eqref{PDE}. For the sake of clarity we repeat the resulting Lagrangian here
\begin{equation}\label{L_CH}
\mathcal L= \ell(\tau)-\frac{2 U}{\dot\ell(\tau)} \,\, ,
\qquad \text{with} \qquad 
\tau=V+\frac{U}{\dot\ell(\tau)^2} \,\, .
\end{equation}
 This approach allowed them to formulate requirements on the causal structure in terms of simple conditions for the function $\ell(\tau)$ and, subsequently \cite{Russo:2025fuc}, provide an alternative description of duality-invariant NLED theories in terms of a new scalar variable $y$ and an associated Legendre transformed function $\Omega(y)$, defined via:
\begin{equation}\label{Legendre_Omega_l}
\ell(\tau)= -\Omega(y)+y\,\Omega_y(y)\,,\qquad \dot{\ell}=y\,,\qquad \tau=\Omega_y\,.
\end{equation}
In terms of these quantities the Lagrangian can be written as:
\begin{equation}\label{L_omega}
    \mathcal{L}=y\,V-\frac U y -\Omega(y)\,,
\end{equation}
and, quite importantly, the second relation in \eqref{L_CH}, which in principle fixes $\tau=\tau(U,V)$ given a choice of $\ell(\tau)$, is now recovered by imposing the vanishing of the variations of $\mathcal{L}$ in \eqref{L_omega} with respect to $y$. In this sense, the new variable $y$ plays the role of a scalar auxiliary field, characterised by an unspecified self-interaction potential $\Omega(y)$, whose explicit form determines the structure of the underlying duality-invariant NLED theory. This picture is highly reminiscent of the IZ NLED in the $\mu$-frame \eqref{L_mu_barmu}, and we will now turn such an intuition into a precise relationship between the two models.

\subsection{RT and $\mu$-frame IZ duality-invariant NLED}\label{subsec:RTvsMuframe}
The starting point in finding the relation between these two theories, is the observation \cite{Ivanov:2002ab} that the $\nu$-frame IZ model \eqref{L_IZ_4d} exhibits SO$(2)$ duality invariance provided that the interaction function only depends on the specific combination $a=\nu\bar{\nu}$ of the auxiliary fields. Despite this restriction, which brings the $\nu$-frame IZ model closer to the duality-invariant RT model, the two theories are still quite different in the nature of their auxiliary fields, which as noted above are not scalars in the $\nu$-frame. The key step at this point is noting that the Legendre transformation \eqref{general_legendre}, leading to the general $\mu$-frame description \eqref{L_mu_barmu}, can also be restricted in such a way that a built-in duality-invariant $\mu$-frame description is obtained. To make this point more evident, let us start by re-expressing the physical fields $\varphi$ and $\bar{\varphi}$, given in equation \eqref{phi-barphi-def} and appearing in the $\nu$-frame Lagrangian \eqref{L_IZ_4d}, in terms of the $S$ and $P$ variables \eqref{S-P-def}:
\begin{equation}
    S=-\frac{1}{2}(\varphi+\bar\varphi)\,,\quad \sqrt{S^2+P^2}=\sqrt{\varphi\bar\varphi}\,.
\end{equation}
Combining this with the relations \eqref{phi_eom}, one can find the following useful rewritings
\begin{equation}\label{eom_alpha_s}
\begin{aligned}
S\deq&-\frac{1}{2}\left((\nu+\bar\nu)(1+aE_a^2)+4aE_a\right)\,,\\
\sqrt{S^2+P^2}\deq&\,\,a^{\frac{1}{2}} \left(1+E_a(\nu+\bar\nu)+E_a^2a\right)\,.
\end{aligned}
\end{equation}
Now we need to introduce the restricted Legendre transform mentioned above, which should naively be performed with respect to the duality-invariant variable $a=\nu\bar\nu$. Such choice is however not the correct one, as we will now show, and it is in fact more convenient to transform with respect to the quantity $\alpha=a^{\frac{1}{2}}$:\footnote{In the following, the reader should not confuse the quantities $\a$ and $\b$ with spinorial indices, as it will be clear from the context.}
\begin{equation}\label{Legendre_E_H_var}
H(\beta)=E-\alpha E_{\alpha} \, ,
\qquad \qquad 
{a^{\frac{1}{2}}}=\alpha \, ,
\qquad\qquad E_{\alpha}=\beta\,.
\end{equation}
From this definition, it follows that $|\beta| =2\sqrt{\mu\bar\mu}$ and the restricted Legendre transform is compatible with the more general one in \eqref{general_legendre}. That is, when $E$ only depends on $a=\nu\bar\nu$ the general double Legendre transform reduces to:
\begin{equation}
\begin{aligned}\label{Legendre_E_H}
H=&E-\nu E_{\nu}-\bar\nu E_{\bar\nu}
\\
=&E-2aE_a
\\
=&E-\alpha E_\alpha \,\, ,
\end{aligned}
\end{equation}
{such that transforming only with respect to $\alpha$ is equivalent to doing the double Legendre transform and then assuming that $E=E(\alpha)$. 
Importantly, due to the extra factor of $2$, this would not have been the case if we had done the Legendre transformation with respect to $a$.
Therefore in this duality-invariant restriction we will only define $H$, the Legendre transform of $E$, as a function of the single variable $\beta$: the latter will turn out to play the same role as the auxiliary field $y$, emerging from the RT description, with the function $H(\beta)$ playing the role of the function $\Omega(y)$, leading to the equivalence of the two descriptions.

To proceed we define $s=\nu+\bar\nu$ and observe that $E_a=\frac{E_{\alpha}}{2\alpha}$, so as to first rewrite the equations \eqref{eom_alpha_s}:
\begin{equation}\label{L_eom_beta_s}
\begin{aligned}
S\deq&-\frac{1}{2}s\left(1+\frac{\beta^2}{4}\right)+\beta H_{\beta}\,,
\\
\sqrt{S^2+P^2}\deq&-H_{\beta}\left(1+\frac{\beta^2}{4}\right)+\frac{\beta s}{2}\, 
 \end{aligned}
 \end{equation}
and subsequently reorganising them into 
\begin{subequations}\label{eom_beta}
\bea
H_{\beta}&\deq&-\frac{\beta S +(1+\frac{\beta^2}{4})\sqrt{S^2+P^2}}{(1-\frac{\beta^2}{4})^2}\,,
\label{eom_beta-a}
\\
s\left(1-\frac{\beta^2}{4}\right)&\deq&-\frac{2\beta \sqrt{S^2+P^2}+2S(1+\frac{\beta^2}{4})}{1-\frac{\beta^2}{4}}\,.
\label{eom_beta-b}
\eea
\end{subequations}
So far, this only shows the existence of a convenient rewriting of the auxiliary field equations of motion in the duality invariant case where $E(\nu,\bar{\nu})=E(\alpha)$ and its corresponding Legendre transform $H(\beta)$. Notice that the expressions involve only the dual variable $\beta$, the Legendre transform function $H$, and its derivative. Hence we expect one could obtain a version of the IZ-$\mu$ frame reviewed in section \ref{subsec:IZ-formalism-4d}, where the auxiliary field is not the complex $\mu$ but the real $\beta$. Hence, the important step in analysing \eqref{eom_beta-b} is actually finding an expression for the $\mu$-frame Lagrangian where these equations should be substituted to replace the auxiliary fields with the physical ones, and, to have that happening with a single real auxiliary field $\beta$. To this aim, it is important to notice that using the auxiliary field equations \eqref{F_eom} and \eqref{phi_eom} to substitute the physical fields into the Lagrangian \eqref{L_IZ_4d} one can obtain an expression that is formally written in terms of the auxiliary fields only (although these depend on the physical fields through the equations of motion). This \emph{on-shell} Lagrangian, written in terms of the Legendre transformed variables \eqref{Legendre_E_H_var}, is:
\begin{equation}\label{L_beta_s}
    \mathcal{L}\deq-\frac{s}{2}\left(1-\frac{\beta^2}{4}\right)+H\,.
\end{equation}
We stress again that the latter expression was obtained from the $\nu$-frame Lagrangian \eqref{L_IZ_4d} by rewriting $\varphi$ and $\bar\varphi$ in terms of the auxiliary fields using \eqref{phi_eom} and then going to the $\mu$-frame \eqref{Legendre_E_H}, under the assumption $E=E(\alpha)$.
Keeping track of this on-shell quantity is convenient to find the expression of an off-shell Lagrangian with the single, real, scalar auxiliary field $\beta$: this is simply the function of $\left(\varphi,\bar\varphi,\beta\right)$ that, in a remarkably simple way, reproduces the equations of motion \eqref{eom_beta-a} when varying $\beta$, and it can be checked that this is true for the following Lagrangian
\begin{equation}\label{L_mu_4D}
    \mathcal{L}=\frac{1+\frac{\beta^2}{4}}{1-\frac{\beta^2}{4}}\,S+\frac{\beta}{1-\frac{\beta^2}{4}}\sqrt{S^2+P^2}+H(\beta)\,.
\end{equation}
Note that since we are trading two auxiliary fields, namely $\nu$ and $\ov \nu $, for a single one $\beta$, the lagrangian \eqref{L_mu_4D} will only give rise to a single auxiliary field equation of motion \eqref{eom_beta-a}, when we vary along $\beta$. This is quite remarkable: we have used the $\nu$-frame equations \eqref{eom_beta-a} and \eqref{eom_beta-b} written in terms of Legendre-transformed quantities to construct the Lagrangian \eqref{L_mu_4D}. However, we also find that upon varying \eqref{L_mu_4D} with respect to $\beta$ we \textit{still} recover \eqref{eom_beta-a} as an equation of motion in this new model with a single scalar auxiliary field. Hence, the idea is to now view \eqref{L_mu_4D} as the \textit{starting} Lagrangian, rather than an on-shell quantity tied to the $\nu$-frame. This will be a recurring theme in all our discussions. 

What we have found is a family of duality-invariant NLED Lagrangians containing a single real scalar auxiliary field $\beta$ characterised by an arbitrary function $H(\beta)$. The physical system described by \eqref{L_mu_4D} is precisely the one described by \eqref{L_IZ_4d} or \eqref{L_mu_barmu}, as can be checked by computing the equations of motion for $F_{\mu\nu}$. Additionally, the expression \eqref{L_mu_4D} coincides with the Lagrangian \eqref{L_omega}, written in terms of a potential (or, so-called, interaction function) $\Omega(y)$ and variables $U$, $V$. The precise correspondence is obtained by expressing the $y$ variable in terms of $\beta$, which eventually also reveals the precise relation between $H(\beta)$ and $\Omega(y)$:
\begin{equation}\label{y(beta)}
y=\frac{1+\frac{\beta}{2}}{1-\frac{\beta}{2}} \,\, ,
\qquad \text{and} \qquad 
H(\beta)=-\Omega\Bigl(\frac{1+\frac{\beta}{2}}{1-\frac{\beta}{2}}\Bigl) \,\, .
\end{equation}
This provides a precise relationship between the IZ auxiliary field formulations and the RT theory of duality-invariant NLED. 

\paragraph{A straightforward sanity check.}
The above result can, for example, be checked against the expression of $H$ for the so-called ModMaxBorn-Infeld model. This was found to correspond to the potential $\Omega(y)=T\left(\frac{y}{2} e^{-\gamma}+\frac{1}{2y}e^{\gamma}-1\right)$ in \cite{Russo:2025fuc} (see, e.g., eq.\,($1.13$) where $\phi=e^y$) and agrees with the result from \cite{Ferko:2023wyi} after applying the change of variable \eqref{y(beta)}.
In \cite{Ferko:2023wyi} the authors had found:
\begin{equation}\label{H_mmb}
\begin{aligned}
H(b)&=\frac{1}{\lambda}\frac{(b-1)+(b+1)\cosh{\gamma}-2\sqrt{b}\sinh{\gamma}}{(b-1)}\,,\quad \frac{\beta}{2}=\sqrt{b}\,,
\\
H\left(\frac{\beta}{2}=\frac{y-1}{y+1}\right)&=-\frac{1}{\lambda}\left[1-\frac{e^{-\gamma} y+e^{\gamma}y^{-1}}{2}\right]\,,
\end{aligned}
\end{equation}
which is precisely $-\Omega(y)$ in \cite{Russo:2025fuc} if we identify $T=\frac{1}{\lambda}$.\footnote{Here $H(b)$ has a minus sign in front of $\gamma$ compared to the expression in \cite{Ferko:2023wyi}. This is due to a typo represented by an incorrect sign in front of the root-$T\bar T$ operator defined in (4.16) of \cite{Ferko:2023wyi}.}

This function is also the solution of a two-parameter flow, corresponding to a $4d$ analogue of the irrelevant $T\bar T$ deformation mixed to the marginal root-$T\bar T$ one. Hence it can be shown that $H$ in \eqref{H_mmb} solves both
\begin{equation}
   \frac{\partial \cL}{\partial \lambda}=\frac18\left(T^{\mu\nu}T_{\mu\nu}-\frac12 (T^{\mu}_{\,\,\,\,\mu})^2\right)
   \qquad \text{and} \qquad 
   \frac{\partial \cL}{\partial \gamma}=\frac12\sqrt{T^{\mu\nu}T_{\mu\nu}-\frac14 (T^{\mu}_{\,\,\,\,\mu})^2}\,,
\end{equation}
when $\cL$ has the form \eqref{L_mu_barmu}. See \cite{Ferko:2023wyi} for a detailed discussion of the relation between duality-invariant electrodynamics and stress-tensor deformations.
Notice that, although they are supposed to describe the same model (when $H$ only depends on $\mu\bar\mu$), it is not a priori obvious that the Lagrangian \eqref{L_mu_4D} gives the same stress-tensor combinations as the one studied in \cite{Ferko:2023wyi}, but this is indeed the case.
The calculation is quite straightforward if one uses the following formulae:
\begin{align}
    T^\mu _{\,\,\,\mu}&=4(\mathcal{L}-S\mathcal{L}_S-P\mathcal{L}_P)\,,\\
    T^{\mu\nu}T_{\mu\nu}&=4(\mathcal{L}-S\mathcal{L}_S-P\mathcal{L}_P)^2+4(S^2+P^2)\mathcal{L}_S^2\,,
\end{align}
which hold for theories where the Lagrangian only depends on $S$ and $P$, as it can be easily checked.
After computing these scalars for the Lagrangian \eqref{L_mu_4D} we can use the equation of motion of the field $\beta$ \eqref{eom_beta-a}. Eventually, we get
\begin{equation}
    T^\mu _{\,\,\,\mu}\deq 4H\,,\qquad\qquad T^{\mu\nu}T_{\mu\nu}\deq 4H^2+4H_\beta^2\left(1-\frac{\beta}{2}\right)^2\,,
\end{equation}
which are equivalent to the expressions found in \cite{Ferko:2023wyi}. Notice also that this is compatible with the results that \cite{Russo:2024xnh} (i.e. eq. (3.7)) found in the CH approach.

Notice that $H$ in \eqref{H_mmb} does not have a limit for $\lambda\to 0$, which is a consequence of the fact that the $\mu$-frame (or $\left(y,\Omega(y)\right)$-frame) cannot describe ModMax electrodynamics \cite{Bandos:2020jsw} (or root-$T\bar T$ deformations). More generally, since $H$ has dimension $[H]=4$ and $\mu$ is dimensionless, the function $H$ will always contain some dimensionful parameter (i.e. $\lambda$ in the previous examples); this is why the $\mu$ frame cannot describe marginal deformations of conformal theories\footnote{With the exception of $H=0$, for which it was shown in the Hamiltonian formalism \cite{Russo:2025fuc} that the theory becomes Bialynicki-Birula electrodynamics \cite{BB:1983}.}. On the other hand, this is not the case for the function $E$ in the $\nu$-frame, where marginal and irrelevant flows are separately well-defined:
\begin{align}\label{MM_BI_nu}
E(\alpha;\gamma)=&~ 2\tanh{\left(\frac{\gamma}{2}\right)}\,\alpha\,,
\\
E(\alpha;\lambda)=&-\frac{3}{8\lambda}\left({}_3F_2\left(-\frac{1}{2},-\frac{1}{4},\frac{1}{4};\frac{1}{3},\frac{2}{3};-\frac{256}{27}(\lambda\alpha)^2\right)-1\right)\,.
    \label{MM_BI_nu-2}
\end{align}
These interaction functions describe respectively ModMax and Born-Infeld theories. Such a feature shows how going from the $\nu$ to the $\mu$-frame amounts to trading a broader range of applicability of $E$ for easier expressions of $H$, in the non-conformal case.

\paragraph{$\nu$-frame and CH correspondence.}\label{subsec:CHvsnuframe}
To conclude our discussion about $4d$ auxiliary field models of electrodynamics, let us show the correspondence between the IZ $\nu$-frame, written in terms of the interaction function $E$, and the Courant-Hilbert approach.
As already mentioned, the $\nu$-frame class of duality-invariant extensions of electrodynamics can be parametrised by restricting $E$ to depend on one single variable $a=\nu\bar\nu$. This is analogous to observing that the duality-invariant condition is solved by the Lagrangian \eqref{L_CH} and it is to be expected that in both models one should be allowed to tune one function of a single variable.

As a quick aside, it is interesting for the analogy with the $\nu$-frame that it is possible to write the Lagrangian \eqref{L_CH} as an auxiliary field Lagrangian, with auxiliary fields $\left(\tau,\lambda\right)$:

\begin{equation}
    \mathcal{L}= \ell(\tau)-\frac{2U}{\dot{\ell}(\tau)}-\lambda\left(\tau-V-\frac{U}{[\dot{\ell}(\tau)]^2}\right)\,,
\end{equation}
where $\lambda$ is a Lagrange multiplier whose equation of motion fixes the condition on $\tau$ in \eqref{L_CH}. This was already mentioned in \cite{Russo:2024llm} and is a way to see the CH solution itself as an auxiliary field theory. Obviously, this agrees with the fact that all the other approaches we have discussed so far, the $\nu$/$\mu$-frames and the RT $y$-scalar auxiliary model, are auxiliary field models. The equation of motion for $\tau$ gives:
\begin{equation}
    \left[\lambda-\dot{\ell}\right]\left(\dot{\ell}^3+2U\ddot{\ell}\right)=0\,,
\end{equation}
which has solution $\dot{\ell}=\lambda$. In \cite{Russo:2024llm}
they argued that the condition $\dot{\ell}^3+2U\ddot{\ell}>0$ ensures strong-field causality of the theory.
In order to match the CH approach with the $\nu$-frame auxiliary model we can try to write the Lagrangian \eqref{L_IZ_4d} in terms of the physical fields and the single auxiliary variable $\alpha$, which we expect to be related to $\tau$. To achieve this we can start from the equations of motion for the auxiliary fields:
\begin{align}\label{V_ab_eom}
    F_{\alpha\beta}&\deq V_{\alpha\beta}(1+E_\nu)=V_{\alpha\beta}\Bigl(1+\frac{E_\alpha \sqrt{\bar\nu}}{2\sqrt{\nu}}\Bigr) \,,\nonumber\\
    \bar F_{\dot\alpha\dot\beta}&\deq \bar V_{\dot\alpha\dot\beta}(1+E_{\bar\nu})=\bar V_{\dot\alpha\dot\beta}\Bigl(1+\frac{E_\alpha \sqrt{\nu}}{2\sqrt{\bar \nu}}\Bigr) \,.
\end{align}
We can manipulate these equations to express the square roots of $\nu$ and $\bar\nu$ as:
\begin{align}\label{nu_of_phi}
    \sqrt{\nu}&\deq\frac{\sqrt{\varphi}-\frac{E_\alpha}{2}\sqrt{\bar\varphi}}{1-(\frac{E_\alpha}{2})^2}\,,\nonumber\\
    \sqrt{\bar\nu}&\deq\frac{\sqrt{\bar\varphi}-\frac{E_\alpha}{2}\sqrt{\varphi}}{1-(\frac{E_\alpha}{2})^2}\,,
\end{align}
meaning that we can use \eqref{V_ab_eom} to write the auxiliary fields $V_{\alpha \beta}$ and $\bar V_{\dot\alpha\dot\beta}$ only in terms of the physical fields ($\varphi$ and $\bar\varphi$) and $\alpha$ (through $E_\alpha$), which allows to rewrite the Lagrangian \eqref{L_AFSM} only in terms of these three variables.\footnote{We can similarly use these equations to show that the equations of motion of the physical fields in the $\nu$-frame and in the $\mu$-frame coincide, which amounts to observing that we are actually looking at the same physical model from two different ``frames''.}
Recalling that $\alpha=\sqrt{\nu\bar\nu}$, one can use the equations \eqref{nu_of_phi} on the Lagrangian \eqref{L_IZ_4d} to find:
\begin{equation}
    \mathcal{L}\deq 2\alpha-\alpha\Bigl(1+\frac{E_\alpha}{2}\Bigr)^2-\frac{(\sqrt{\varphi}+\sqrt{\bar\varphi})^2}{2(1+\frac{E_\alpha}{2})}\Bigl(1-\frac{E_\alpha}{2}\Bigr)+ E\,.
\end{equation}
In this expression $\alpha$ implicitly depends on $\varphi$, $\bar\varphi$, and the choice of $E$ through \eqref{nu_of_phi}.
This Lagrangian has exactly the same form as \eqref{L_CH} once one notices that $U=\frac{1}{4}(\sqrt{\varphi}+\sqrt{\bar\varphi})^2$. Hence, comparing with \eqref{L_CH}, we can identify:
\begin{equation}\label{l_vs_E}
    \ell=2\alpha-\alpha\Bigl(1+\frac{E_\alpha}{2}\Bigr)^2+E
    \qquad \text{and} \qquad \dot\ell=\frac{1+\frac{E_\alpha}{2}}{1-\frac{E_\alpha}{2}}\,.
\end{equation}
This is the relation between the interaction function $E$ and the CH function $\ell$, but to be able to use it we need to know how $\tau $ is defined in terms of $\alpha$.
In order to understand how $\alpha$ and $\tau$ are related we must exploit the above relationship between the RT and $\mu$-frame IZ approaches. Firstly, the connection between the CH approach and the $y$-auxiliary field Lagrangian is that $\Omega(y)$ can be defined from $\ell$ via a Legendre transformation.
Then, since we have just shown that $H=-\Omega$, this means that both $E$ and $\ell$ can be obtained as Legendre anti-transformations of $\Omega$, although with respect to different variables. Recalling \eqref{Legendre_E_H} we know that $E$ is the Legendre transformation of $H$ with respect to the variable $\beta$. On the other hand, the Legendre transform connecting $\Omega$ and $\ell$ is with respect to $y$. Indeed, by using the definition of $\ell$ from \cite{Russo:2025fuc},
\begin{equation}
\ell= -\Omega+y\,\Omega_y\,,
\qquad \qquad  
\dot{\ell}=y\,,
\qquad \qquad
\tau=\Omega_y\,,
\end{equation}
and recalling that $y=\frac{1+\frac\beta2}{1-\frac\beta2}$, one obtains:
\begin{equation}
    \dot\ell= \frac{1+\frac{\beta}{2}}{1-\frac{\beta}{2}} \qquad \text{and} \qquad  \tau=\frac{\partial \beta}{\partial y} \Omega_{\beta}\,.
\end{equation}
Since $E_\alpha=\beta$, this expression for $\dot\ell$ agrees with the one we have found in \eqref{l_vs_E}.
Using $y=\frac{1+\frac\beta2}{1-\frac\beta2}$ we can now express the derivative $\frac{\partial\beta}{\partial y}$ in terms of $\beta$, obtaining expressions depending on $\beta$ and $\Omega_\beta$ only:
\begin{equation}
    \frac{\partial\beta}{\partial y}=\frac{\partial}{\partial y}\frac{2(y-1)}{1+y}=\frac{4}{(1+y)^2}=\Bigl(1-\frac\beta2\Bigr)^2 \,\, , \qquad\qquad \tau=\Omega_\beta\Bigl(1-\frac\beta2\Bigr)^2\,.
\end{equation}
At this point one can rewrite everything in terms of $E$ and $\alpha$, using \eqref{Legendre_E_H} and recalling that $H=-\Omega$. Finally we get:
\begin{equation}\label{l_E}
    \dot\ell=\frac{1+\frac{E_\alpha}{2}}{1-\frac{E_\alpha}{2}}
    \qquad \text{and} \qquad
    \tau=\alpha\Bigl(1-\frac{E_\alpha}{2}\Bigr)^2\,.
\end{equation}
This provides the relation between $\tau $ and $\alpha$ and looking back at \eqref{l_vs_E} it is possible to check that equations \eqref{l_E} can be directly integrated, leading to the expression of $\ell$ in the first equation of \eqref{l_vs_E}.

Having shown this last identification completes the derivation of the correspondence between the IZ model, restricted to self-dual theories, and the model that has been recently studied by Russo and Townsend. To summarise the main takeaways: we have shown that the IZ model in the $\mu$-frame is equivalent to the scalar auxiliary approach of \cite{Russo:2025fuc}. The two functions appearing in the Lagrangians for these two models satisfy $H=-\Omega$. These two models are naturally expressed respectively with the auxiliary scalars $\beta$ and $y$, related by $y=\frac{1+\frac\beta2}{1-\frac{\beta }{2}}$.
Moreover, knowing that:
\begin{itemize}
    \item the $ \nu$-frame is connected to the $\mu$-frame by a Legendre transformation of the field $\beta$\,,
    \item the CH solution is connected to the $y$-auxiliary model of RT by a Legendre transformation of the field $y$\,, 
\end{itemize}
we argued how the $\nu$-frame and CH can also be expressed one in terms of the other. In \eqref{l_vs_E} and \eqref{l_E} we have derived the expression for the pair $(\ell,\tau)$ of the CH solution in terms of the $\nu$-frame interaction function $E$ and the field $\alpha$. This concludes the derivation of the equivalence of all of these different models. 
In the remainder of this work, the analysis carried out here will be crucial to study a special class of models in two spacetime dimensions.

\section{Two-dimensional sigma models in the $\mu$-frame}\label{sec:mu-frame-2d} \label{2D_muframe} Deformations of two-dimensional integrable sigma models using non-scalar and Lie-algebra-valued vector auxiliary fields were first introduced in \cite{Ferko:2024ali} for Principal Chiral Models (PCMs), and have since then been used to study integrable deformations of various other classes of sigma models \cite{Bielli:2024ach, Bielli:2025uiv, Bielli:2024fnp, Bielli:2024oif, Cesaro:2024ipq,Fukushima:2024nxm,Ferko:2025bhv}. The main underlying strategy has been inspired from the $\nu$-frame auxiliary field electrodynamics of Ivanov and Zupnik studied in the previous section, and hence relies on a simple coupling of physical fields to non-dynamical fields, in turn characterised by a choice of self-interaction function. 

Starting from the model in \cite{Ferko:2024ali}, we will now show that, along the lines of what was observed in $4d$ electrodynamics, it is possible to relate the $\nu$-frame-deformed PCM to the Courant-Hilbert solution \cite{Babaei-Aghbolagh:2025uoz}, and therefore to the $y$-auxiliary scalar model, via the $\mu$-frame. Later on we will also show how some of the results can be generalised, using a similar logic, to other classes of sigma models that have been studied as extensions of this $\nu$-frame auxiliary field formalism \cite{Bielli:2025uiv}. 

To begin, let us briefly recall a few relevant details about PCMs and their coupling to auxiliary fields. The undeformed models are $2d$ sigma models on a Lie group, whose fundamental fields are group elements $g(\sigma,\tau)\in$ G, depending on the $2d$ worldsheet coordinates. Their Lagrangian, invariant under the left and right action of the group, can be defined in terms of the pullback to the worldsheet of the left-invariant Maurer-Cartan form
\begin{equation}\label{L_PCM}    \mathcal{L}_{\text{PCM}}=\frac{1}{2}\eta^{\alpha\beta} \, \mathrm{tr}(j_{\alpha}j_{\beta})=-\frac{1}{2}\mathrm{tr}(j_+j_-) \,\, 
\qquad \text{with} \qquad 
j=g^{-1}\mathrm{d}g \in \mathfrak{g}=\text{Lie(G)} \,\, ,
\end{equation}
where $\eta^{\alpha\beta}$ denotes the flat inverse metric on the worldsheet and in the second identity we exploited lightcone coordinates following the conventions in \cite{Ferko:2024ali}. Following again the original paper, the above theory can be coupled to auxiliary fields in the following way:
\begin{equation}\label{L_AFSM}
\mathcal{L}=\frac{1}{2} \mathrm{tr}(j_+j_-)+\mathrm{tr}(v_+j_-)+\mathrm{tr}(j_+v_-)+ \mathrm{tr}(v_+v_-)+ E(v_+,v_-)\,,
\end{equation}
where the interaction function $E$ depends in general on scalar combinations of $v_+$ and $v_-$, that could in principle be written in terms of traces of arbitrary products of $v_\pm$. When the interaction function is restricted to depend on the specific combination $\nu_2=\mathrm{tr}(v_+^2)\mathrm{tr}(v_-^2)$, the model has been shown to describe $T\bar T$-like stress tensor deformations \cite{Ferko:2024ali}. It is also possible to check that integrability is guaranteed provided that $E$ only depends on traces $\mathrm{tr}(v_\pm^n)$, see for example \cite{Bielli:2025uiv}, which have been shown to induce higher-spin integrable deformations of the Smirnov-Zamolodchikov type \cite{Smirnov:2016lqw}. While initially only single combinations of the form $\mathrm{tr}(v_+^n)\mathrm{tr}(v_-^n)$ had been taken into account \cite{Bielli:2024ach,Bielli:2024fnp,Bielli:2024oif}, it was later shown in \cite{Bielli:2025uiv} that in order to generate consistent integrable higher-spin flows it is necessary to assume $E$ to be a function of any Lorentz invariant combination of the traces $\mathrm{tr}(v_+^m)$, $\mathrm{tr}(v_-^n)$, with $m$ and $n$ in principle arbitrary integers. 

Here we will restrict our attention to the case where $E$ is only a function of traces containing up to two auxiliary fields $v_\pm$, which is when one would expect a $\mu$-frame to describe $T\bar{T}$-like deformations.
As an extension of \cite{Ferko:2024ali}, we will show that integrability can be ensured not only by the dependence of $E$ on $\nu_2$, but also by a suitable inclusion of the simplest non-chiral single-trace operator 
$\tr(v_+v_-)$.
The logic will be analogous to the one used in $4d$ electrodynamics, and for this reason, we start once again by studying the auxiliary field equations of motion. The interaction function  $E$ is assumed to depend on the following two variables only
\begin{equation}\label{2d-p-nu-definition}
    \nu = \sqrt{\mathrm{tr}(v_+^2)\mathrm{tr}(v_-^2)}\,,
    \qquad\qquad
    p= \mathrm{tr}(v_+v_-)\,,
\end{equation}
hence leading to equations of motion for the fields $v_+$ and $v_-$ of the form
\begin{equation}\label{eom_p_nu}
\begin{aligned}
j_++v_+(1+E_p)+v_- \sqrt{\frac{\nu_{+2}}{\nu_{-2}} } E_{\nu}\deq&~0\,,
\\
j_-+v_-(1+E_p)+v_+ \sqrt{\frac{\nu_{-2}}{\nu_{+2}} }E_{\nu}\deq&~0\,,
\end{aligned}
\end{equation}
where $E_p=\frac{\partial E}{\partial p}$ and $E_\nu=\frac{\partial E}{\partial \nu}$ and $\nu_{\pm2}=\mathrm{tr}(v_\pm^2)$. Also in this setting, the symbol ``$\deq$'' is used to denote an identity which holds true when the auxiliary field equations of motion \eqref{eom_p_nu} are satisfied.
As in the $4d$ case, it is convenient to use such equations to express the Lagrangian purely in terms of the auxiliary fields: using \eqref{eom_p_nu}, one can find that
\begin{align}
    \mathrm{tr}(j_+j_-)\deq&~ p\left[(1+E_p)^2+E_{\nu}^2\right] +2 E_{\nu} \nu (1+E_p)\,,\\
    \mathrm{tr}(j_{\pm}v_{\mp})\deq&-p (1+E_p) -E_{\nu} \, \, ,
\end{align}
and substituting these relations into the Lagrangian \eqref{L_AFSM} leads to
\begin{align} \label{L_p_nu}
    \mathcal{L}\deq\frac{1}{2}p\left(-1-2E_p+E_p^2+E_{\nu}^2\right) +E_{\nu}E_p\nu  - E_{\nu}\nu  +&E(p,\nu) \,.
\end{align}
This is the same step that we carried out for electrodynamics in \eqref{L_beta_s}, with the slight difference that $E$ is here allowed to depend on two variables.
From this expression, it appears that the auxiliary field structure in two dimensions is slightly different from the one in four dimensions and if one wants to study single variable deformations of the PCM it is possible to simply require $E$ to depend on either $\nu$ or $p$ only. To begin, in the next subsection, we will briefly focus on the former case, which corresponds to the $\mu$-frame of the two-dimensional family of integrable models studied in \cite{Ferko:2024ali}. We will then show how this formalism can be expanded to describe the combined dependence on $\nu$ and $p$, also extending the analysis of the underlying integrable structure in the presence of the latter extra variable in section \ref{s:int}.

\subsection{$\mu$-frame for $E(\nu)$ - the case of AFSM}

Restricting ourselves to the assumption $E_p=0$, the on-shell Lagrangian \eqref{L_p_nu} reduces to
\begin{equation}\label{on-shell-Lag-mu-frame-pcn}
    \mathcal{L}\deq\frac{1}{2}p\left(-1+E_{\nu}^2\right)  - E_{\nu}\nu \ +E(\nu) \,,
\end{equation}
and from the equations of motion \eqref{eom_p_nu} one obtains
\begin{equation}\label{traces-mu-frame-pcm}
\begin{aligned}
\mathrm{tr}(j_+j_-)\deq& \,p \left[1+E_{\nu}^2\right] +2 E_{\nu} \nu\,,
\\
\mathrm{tr}(j_+^2) \mathrm{tr}(j_-^2)\deq&\left[(1+E_{\nu}^2)\nu+2 E_{\nu}p\right]^2\,.
\end{aligned}
\end{equation}
Performing a Legendre-transform with respect to $\nu$
\begin{equation}
    H(\mu)=E-\nu E_{\nu}\,,
    \qquad \qquad \mu=E_\nu\,,
\end{equation}
and using \eqref{traces-mu-frame-pcm} one can then rewrite the on-shell Lagrangian \eqref{on-shell-Lag-mu-frame-pcn} as:
\begin{equation}\label{L_p_mu}
    \mathcal{L}\deq-\frac{1}{2}p\left(1-\mu^2\right) +H(\mu)\,,
\end{equation}
and similarly, the relations \eqref{traces-mu-frame-pcm} become
\begin{equation}\label{trace_j_relations}
\begin{aligned} 
\mathrm{tr}(j_+j_-)\deq& \,p \left[1+\mu^2\right] -2 \mu H_{\mu}\,,
\\
\mathrm{tr}(j_+^2) \mathrm{tr}(j_-^2)\deq&\left[(1+\mu^2)(-H_{\mu})+2 \mu p\right]^2\,.
\end{aligned}
\end{equation}
At this point, differently from what happens in \cite{Ivanov:2003uj,Ferko:2023wyi}, the on-shell quantity $p$ appears in the Lagrangian instead of $H_{\mu}$. To find the expression for the off-shell Lagrangian reproducing these equations of motion it is convenient to look at \eqref{trace_j_relations}:
we can multiply and sum up the two equations, obtaining convenient relations
    \begin{subequations}
    \begin{equation}\label{trace_j_relations1}
      p\deq -\frac{2\mu \sqrt{\mathrm{tr}(j_+^2)\mathrm{tr}(j_-^2)}-(1+\mu^2)\mathrm{tr}(j_+j_-)}{(1-\mu^2)^2}\,,  
    \end{equation}
    \begin{equation}\label{trace_j_relations2}
       H_{\mu}\deq -\frac{(1+\mu^2)\sqrt{\mathrm{tr}(j_+^2)\mathrm{tr}(j_-^2)}-2\mu \,\mathrm{tr}(j_+j_-)}{(1-\mu^2)^2}\,\,. 
    \end{equation}
    \end{subequations}
From the intuition developed in the previous section we can then write down a Lagrangian and check that the equation of motion \eqref{trace_j_relations2} is correctly reproduced when varying 
\begin{equation}\label{L_j_mu}
    \mathcal{L}=\frac{1}{2}\frac{2\mu \sqrt{\mathrm{tr}(j_+^2)\mathrm{tr}(j_-^2)}-(1+\mu^2)\mathrm{tr}(j_+j_-)}{(1-\mu^2)}+H(\mu)\,
\end{equation}
along $\mu$. In particular, the equation of motion can be found by simply taking a derivative of 
\eqref{L_j_mu} with respect to $\mu$ and noting that setting this to zero imposes exactly the relation \eqref{trace_j_relations2} for $H_\mu$. The equation \eqref{trace_j_relations1} is not satisfied when obtaining the equations of motion varying \eqref{L_j_mu}.
We hence conclude that \eqref{L_j_mu} is the $\mu$-frame Lagrangian for the auxiliary field PCM in the $\nu$-frame constructed in \cite{Ferko:2024ali}, in the case when $E=E(\nu)$ is a function only of the real variable $\nu$. 
As one would have expected from the $4d$ setting, the new model only exhibits one auxiliary field and is, at first sight, very different from the model we started from. We explicitly checked in appendix \ref{app_a} that the equations of motion for the $\nu$-frame model correctly reduce to the ones for the $\mu$-frame, confirming that these are two descriptions of the same physical system.

We can also use this model to describe stress tensor deformations of the PCM as done in previous work, see e.g. \cite{Ferko:2024ali,Bielli:2024khq,Bielli:2024ach,Babaei-Aghbolagh:2025uoz}. We can define the $T\bar T$ and the root-$T\bar T$ operators as:
\begin{equation}
\begin{aligned}
\mathcal{O}_{T\bar T}=&~T_{\mu\nu}T^{\mu\nu}-(T^{\mu}_{\,\,\,\mu})^2\,,
\\
\mathcal{R}_{\sqrt{T\bar T}}=&~\frac{1}{\sqrt{2}}\sqrt{T_{\mu\nu}T^{\mu\nu}-\frac{1}{2}(T^{\mu}_{\,\,\,\mu})^2}\,.
\end{aligned}
\end{equation}
Translating them to the $\mu$-frame, as shown in appendix \ref{app_b}, one finds 
\begin{equation}\label{flows-mu-frame}
\begin{aligned}
\mathcal{O}_{T\bar T}=&~\frac{1}{2}(H_\mu)^2(1-\mu^2)^2-2H^2,
\\
\mathcal{O}_{\sqrt{T\bar T}}=&~\frac{1}{\sqrt{2}}\sqrt{\frac{1}{2}(H_\mu)^2(1-\mu^2)^2}\,.
\end{aligned}
\end{equation}
It can also be checked that these are the same expressions that one would obtain by simply Legendre-transforming the operators to the $\mu$-frame as in appendix \ref{app_b}. Finally, imposing that the Lagrangian should satisfy both flows gives rise to the following solution
\begin{equation}\label{H_lambda_gamma}
\frac{\partial \mathcal{L}}{\partial \lambda\,,\gamma}=\mathcal{O}_{T\bar T\,,\sqrt{T\bar T}}
\quad \Rightarrow \quad     H(\mu)^{(\gamma,\lambda)}=\frac{1}{4\lambda}\frac{(\mu^2-1)+(\mu^2+1)\cosh{\gamma}-2\mu\sinh{\gamma}}{(\mu^2-1)}\,.
\end{equation}
In analogy with the $4d$ case, these results are again compatible with the analysis of stress-tensor flows for the $2d$ analogue of the $y$-auxiliary model of Russo and Townsend \cite{Babaei-Aghbolagh:2025uoz}, such that one can again identify the $\mu$-frame with this recent theory by simply choosing:
\begin{equation}\label{y(mu)}
y=\frac{1+\mu}{1-\mu}
\qquad \text{and} \qquad
H(\mu)=-\Omega(y) \,\, .
\end{equation}

It should also be noted that the two flows \eqref{flows-mu-frame} coincide, in the $\nu$-frame, with the analogous $4d$ flows reported in \eqref{MM_BI_nu} and \eqref{MM_BI_nu-2}.
To complete the mapping between all these different descriptions, it is then possible to find a relation between $E$ and $\ell$ analogous to the one found for NLED in \eqref{l_vs_E}:
\begin{equation}
    \dot \ell =\frac{1+E_\nu}{1-E_\nu}\,,
    \qquad\qquad
    \tau= \frac{\nu}{2}(1-E_\nu)^2\,, \,
\end{equation}
which, after integration, or using the equations of motion on \eqref{L_p_nu}, lead to:
\begin{equation}
    \ell=\frac{\nu}{2}(1-E_\nu^2)-\nu E_\nu+E\,.
\end{equation}
To conclude this section, it should be stressed that the \emph{$\mu$-frame} not only represents a rephrasing of the \emph{$\nu$-frame}, but also allows to describe models that could not be defined from the latter point of view. The reason can be found in the fact that these two models are connected by a Legendre transformation between $E$ and $H$: clearly, this is well-defined for convex functions and hence $H=0$, which is the only possible choice of $H$ preserving (classical) conformal invariance, has no analogue in the $\nu$-frame.
This choice was already shown in four dimensions to correspond to the Bialynicki-Birula (BB) electrodynamics \cite{BB:1983}, the only other conformal and $SO(2)$-duality invariant theory of electrodynamics together with ModMax theory \cite{Bandos:2020jsw}. It is well known that BB electrodynamics does not have a Lagrangian description, since imposing $H=0$ in the auxiliary field electrodynamics model implies that the auxiliary field equations of motion enforce the Lagrangian itself to be zero.
Here, it is interesting to notice that in $2d$ such complication does not arise: starting from \eqref{L_j_mu} and setting $H=0$ one finds that the auxiliary field $\mu$ can actually be integrated out, leading to the Lagrangian:
\begin{equation}\label{mu-frame-2d-H=0-Lagrangian}
    \mathcal{L}=\frac{1}{2}\sqrt{\mathrm{tr}(j_+j_-)^2-\mathrm{tr}(j_+j_+)\mathrm{tr}(j_-j_-)}\,.
\end{equation}
Hence, quite differently, in this case, the $H=0$ theory is at least defined by a Lagrangian, albeit a non-analytic one.
As in the $4d$ case, we can again interpret this theory as some sort of strong field limit, that is $\lambda \to \infty$ limit, of the $T\bar{T}$-deformed PCM. This can also be seen from the expression of the interaction function for the $T\bar T$ and root-$T\bar T$ mixed flows \eqref{H_lambda_gamma}, which goes to zero in the $\lambda \to \infty$ limit.
Finally, it should be noted that the Lagrangian \eqref{mu-frame-2d-H=0-Lagrangian} goes to zero only in the extreme case of a single boson. The $(N>1)$-boson case or, more generally, the non-linear sigma model case, are non-zero and can be compared to the $\lambda \to \infty$ limit of the results found in \cite{Cavaglia:2016oda, Bonelli:2018kik}.

\subsection{$(\mu, \rho)$-frame for $E(\nu,p)$}\label{s:murhofr}
In this subsection, we extend the above reasoning by allowing the $\nu$-frame interaction function to depend not only on the combination $\nu$ of auxiliary fields, but also on $p$ in \eqref{2d-p-nu-definition}. Taking this into account requires performing a double Legendre transform leading to a new function $H(\mu\,,\rho)$ characterising a new combined $(\mu,\rho)$-frame:
\begin{equation}\label{Legendre_mu_mutilde}
H=E-\nu E_\nu-pE_p\,,
\qquad \qquad
\mu=E_\nu\,,
\qquad \qquad
\rho=E_p\,.
\end{equation}
Analogously, we will also have $\nu=-H_{\mu}$ and $p=-H_{\rho}$, as per the definition of the Legendre transformation,  and our goal will be to find a Lagrangian $\mathcal{L}(j_\pm,\mu,\rho)$ describing a system equivalent to the AFSM \eqref{L_AFSM} for $E=E(\nu,p)$, such that the model \eqref{L_j_mu} is recovered in the limiting case of $\rho \rightarrow0$. 

To construct the $(\mu,\rho)$-model we proceed as before: from the auxiliary field equations of motion \eqref{eom_p_nu} one finds
\begin{equation}
\begin{aligned}
    \mathrm{tr}(j_+j_-)\deq&~ p \left[(1+E_p)^2+E_{\nu}^2\right] +2 E_{\nu} \nu(1+E_p)\,,
    \\
    \mathrm{tr}(j_+^2) \mathrm{tr}(j_-^2)\deq&\left[\bigl((1+E_p)^2+E_{\nu}^2\bigr)\nu+2 E_{\nu}p(1+E_p)\right]^2\, ,
\end{aligned}
\end{equation}
in terms of which the on-shell Lagrangian takes the form \eqref{L_p_nu}. One can now take the Legendre transform \eqref{Legendre_mu_mutilde} and rearrange the equations of motion of the auxiliaries as:
\begin{equation}\label{trace_j_relations3}
\begin{aligned}
H_{\rho}\deq&~ \frac{2\mu(1+\rho) \sqrt{\mathrm{tr}(j_+^2)\mathrm{tr}(j_-^2)}-((1+\rho)^2+\mu^2)\mathrm{tr}(j_+j_-)}{((1+\rho)^2-\mu^2)^2}\,,
\\
H_{\mu}\deq& -\frac{((1+\rho)^2+\mu^2)\sqrt{\mathrm{tr}(j_+^2)\mathrm{tr}(j_-^2)}-2\mu (1+\rho)\,\mathrm{tr}(j_+j_-)}{((1+\rho)^2-\mu^2)^2}\,\, .
\end{aligned}
\end{equation}
Similarly, this operation brings the Lagrangian to the form
\begin{equation}
    \mathcal{L}\deq-\frac{H_{\rho}}{2}(-1+\rho^2+\mu^2)- \rho\mu H_\mu  +H\, ,
\end{equation}
and after substituting the relations \eqref{trace_j_relations3} one can interpret the resulting expression as a new off-shell Lagrangian, checking that the equations of motion \eqref{trace_j_relations3} are recovered when we treat $\mu$ and $\rho$ as independent auxiliary fields.
It is easy to verify that such an operation leads to the following $(\mu,\rho)$-frame Lagrangian
\begin{equation}\label{L_mu_rho}
\mathcal{L}=\frac{2\mu\sqrt{\mathrm{tr}(j_+^2)\mathrm{tr}(j_-^2)}+(-1+\rho^2-\mu^2)\mathrm{tr}(j_+j_-)}{2(1+\rho-\mu)(1+\rho+\mu)}+H(\mu,\rho)\,.
\end{equation}
As expected, the simpler expression \eqref{L_j_mu} is correctly recovered when $\rho=0$. In order to confirm that this model is physically equivalent to the original $\nu$-frame, we also checked that the physical fields equations of motion arising from \eqref{L_mu_rho} coincide with the ones from \eqref{L_AFSM}, reporting the details of the calculation in appendix \ref{app_a}. As explained in the appendix, the $\mu$-frame equations of motion are recovered by imposing the following condition on the $v_\pm$ in the $\nu$-frame:
\begin{equation}
    v_{\pm}\deq\frac{-j_{\pm} (1+E_p)+j_{\mp}\mathrm{tr}(j_{\pm}^2)^{\frac{1}{2}}\mathrm{tr}(j_{\mp}^2)^{-\frac{1}{2}}E_\nu}{(1+E_p)^2-E_\nu^2}\,.
\end{equation}
This equation can hence be used, together with the identification \eqref{Legendre_mu_mutilde} for $E_\nu$ and $E_p$, to pass from expressions ``in the $\nu$-frame'' to expressions ``in the $\mu$-frame''. We should finally stress that also in this case the new Lagrangian \eqref{L_mu_rho} exhibits an auxiliary field structure which is significantly simpler, in comparison with the starting Lagrangian \eqref{L_AFSM}, since non-scalar and Lie-algebra-valued auxiliary fields have been traded for purely scalar fields. While this simplification in the auxiliary sector is partly compensated by a complication of the physical one, we will now show how certain features will turn out to be simpler in this new picture, by studying the integrable structure of the $(\mu,\rho)$-frame theory.

\section{Integrability in the two-dimensional $\mu$-frame }\label{s:int}
As already mentioned, the auxiliary field PCM \eqref{L_AFSM} in the $\nu$-frame was shown to be integrable in \cite{Ferko:2024ali} under the assumption of an interaction function $E$ depending only on the variable $\nu_2=\mathrm{tr}(v_+^2)\mathrm{tr}(v_-^2)$.\footnote{In fact, a broader sufficient condition for integrability is that the interaction function might depend on any combination of the variables $\nu_{\pm n}:=\mathrm{tr}(v_{\pm}^n)$  \cite{Bielli:2024ach,Bielli:2025uiv}, as can be checked by demanding that \eqref{important_commutators} hold true on-shell. The case of higher-spin deformations, associated with the presence of variables having $n>2$, will not be considered in this work.}
Here we will turn our attention to the integrable structure underlying the $\mu$-frame two-dimensional theories constructed in section \ref{sec:mu-frame-2d}.
We will start by showing how the equations of motion can be encoded in the flatness condition of a Lax connection, a property which ensures the existence of an infinite set of conserved charges \cite{Torrielli:2016ufi, Driezen:2021cpd, Lacroix:2018njs} and is sometimes referred to as weak integrability, first for the simpler $\mu$-frame description of AFSM and successively for the combined $(\mu,\rho)$-frame. In the last part of this section, we will also discuss the Poisson structure of $\mu$-frame theories, describing how the spatial components of the Lax satisfy the Maillet structure, which ensures involution of the charges, a condition sometimes referred to in the literature as strong integrability. 

\paragraph{$\nu$-frame integrability and Courant-Hilbert}
To begin, let us briefly review the observed similarity between $2d$ integrability and the $4d$ Courant Hilbert solution underlying the $\nu$-frame AFSM construction in \cite{Ferko:2024ali}.
 To highlight this connection, let us take a step back from the auxiliary fields and consider a generic Lagrangian $\mathcal{L}=\mathcal{L}(x_1,x_2)$, containing no auxiliary fields and with:
\begin{equation}\label{frakJ_def}
    x_1=-\mathrm{tr}(j_+j_-)\,,
   \qquad\qquad 
    x_2=\frac{1}{2}\left[\mathrm{tr}(j_+^2)\mathrm{tr}(j_-^2)+\mathrm{tr}(j_+j_-)^2\right]\,.
\end{equation}
For these Lagrangians, one can define the conserved current corresponding to the right group multiplication symmetry of the model:
\begin{equation}\label{current_x12}
\frak{J}_{\alpha} =2 \frac{\partial \mathcal{L}}{\partial x_1}j_{\alpha} + 4 \frac{\partial \mathcal{L}}{\partial x_2}j^{\beta}\mathrm{tr}(j_{\alpha}j_{\beta})\,, 
\qquad \qquad
\partial^\alpha\frak{J}_\alpha\approx 0\,.
\end{equation}
It was shown in \cite{Borsato:2022tmu} that it is possible to define the components of the Lax connection as:
\begin{equation}\label{Lax_general}
\frak{L}_{\pm}=\frac{j_{\pm}\pm z\frak{J}_{\pm}}{1-z^2}\, .
\end{equation}
As desired, this Lax obeys the flatness condition when one uses the equations of motion:
\begin{equation}
\mathrm{d}_{\frak{L}}\frak{L}=\partial_+\frak{L}_--\partial_-\frak{L}_++\left[\frak{L}_+,\frak{L}_-\right]\approx0\,, 
\end{equation}
where the symbol ``$\approx$'' is used to denote equality when the equations of motion of physical fields are imposed.
For the above flatness equation to hold it is necessary that the following two relations are satisfied:
\begin{equation}\label{important_commutators}
\begin{aligned}
\left[\frak{J}_+,\frak{J}_-\right]&=\left[j_+,j_-\right]\,,
\\
\left[\frak{J}_+,j_-\right]&=\left[j_+,\frak{J}_- \right]\, .
\end{aligned}
\end{equation}
This can be easily verified explicitly:
\begin{align}\label{flatness}
\mathrm{d}_{\frak{L}}\frak{L}=& \, \partial_+\frak{L}_--\partial_-\frak{L}_++\left[\frak{L}_+,\frak{L}_-\right]
    \notag \\
    =& \,\frac{1}{1\!-\!z^2}\!\left(\!\partial_{+}j_{-}-\partial_{-}j_{+} \!-\! z\Bigl(\!\partial_{+}\frak{J}_{-}+\partial_{-}\frak{J}_{+}\!\Bigr)\!+\!\frac{\left[j_+,j_-\right]\!-\!z^2\left[\frak{J}_+,\frak{J}_-\right]\!+\!z\left(\left[\frak{J}_+,j_-\right]\!-\!\left[j_+,\frak{J}_- \right]\right)}{1-z^2}\!\right)
    \notag \\
    \deq& \,\frac{1}{1\!-\!z^2}\Bigl(\!\partial_{+}j_{-}-\partial_{-}j_{+}+[j_+,j_-]-z\left(\partial_{+}\frak{J}_{-}+\partial_{-}\frak{J}_{+}\right)\!\Bigr) \approx \, 0 \,\, ,
\end{align}
where in the second line we have imposed the commutation relations \eqref{important_commutators} and at the end we are left with the divergence of $\mathfrak{J}$, which vanishes on-shell, and the flatness of $j$, which is automatically ensured by the Maurer-Cartan equation.

Hence, for a Lagrangian $\mathcal{L}=\mathcal{L}(x_{1},x_{2})$ the associated Lax \eqref{Lax_general} enjoys the desired on-shell flatness condition provided that \eqref{important_commutators} hold true: from the definition \eqref{current_x12} of $\mathfrak{J}$ one can then recognise that imposing \eqref{important_commutators} amounts to solving a partial differential equation involving $\frac{\partial\mathcal{L}}{\partial x_1}$ and $\frac{\partial\mathcal{L}}{\partial x_2}$, which upon introducing the new variables:
\begin{equation}
    \tilde{U}=\frac{1}{4}\left(\sqrt{2x_2-x_1^2}-x_1\right)
    \qquad \text{and} \qquad 
    \tilde{V}=\frac{1}{4}\left(\sqrt{2x_2-x_1^2}+x_1\right)\,,
\end{equation}
takes the form
\begin{equation} \label{LaxPDE}
    \mathcal{L}_{\tilde{U}}\mathcal{L}_{\tilde{V}}=-1\,.
\end{equation}
This is analogous to the self-duality condition \eqref{PDE_intro} mentioned in section \ref{sec:intro}, where the generic $4d$ electrodynamics Lagrangian function of $(S,P)$ has been replaced by a $2d$ sigma model Lagrangian depending on $(x_1,x_2)$.
This fact was already observed in \cite{Borsato:2022tmu, Ferko:2023wyi, Babaei-Aghbolagh:2025uoz} and independently of the dimensionality, the Courant-Hilbert approach is once again the most general solution to equation \eqref{LaxPDE}. 
In the same spirit, the definition of the AFSM \eqref{L_AFSM} in \cite{Ferko:2024ali} allowed them to define the conserved current:
\begin{equation}\label{frac_j_nu_conservation}
    \frak{J}_{\alpha}=-(j_\alpha+2v_{\alpha})\,,
    \qquad\qquad 
     \partial^{\alpha}\frak{J}_{\alpha}\dapp0\,.
\end{equation}
The authors of \cite{Ferko:2024ali} have shown that one can use this current in place of the $\frak{J}$ defined from \eqref{current_x12} inside a Lax connection of the same form as \eqref{Lax_general}. Crucially, the structure of the auxiliary fields $v_\pm$ implies that \eqref{important_commutators} is satisfied when the auxiliary fields are on-shell. Hence the flatness of the Lax connection of the AFSM in the $\nu$-frame follows from the previous calculation. In practice, the power of the auxiliary fields in the $\nu$-frame is that the interaction function $E$ does not enter the explicit definition of the Lax. One can then express this ``auxiliary'' Lax connection, by specifying an interaction function $E$ and setting $v_\pm$ on-shell, in terms of a more standard Lax connection built out of the physical fields.

\subsection{Lax in the $\mu$-frame - the case of AFSM}
Having briefly reviewed the connection between the Courant-Hilbert solution and $2d$ integrability, we can now proceed in constructing the Lax connection for the auxiliary field model in the $\mu$-frame. In section \ref{sec:mu-frame-2d} we have shown that the $\mu$-frame coincides with the $\left(y,\,\Omega(y)\right)$ auxiliary field model \cite{Russo:2025fuc}: this solves \eqref{LaxPDE}, since it is equivalent to the Courant-Hilbert solution, and therefore we expect the Lax to obey again the general structure \eqref{Lax_general}.
Starting from the Lagrangian \eqref{L_j_mu} we can hence compute the current \eqref{current_x12}:
\begin{equation}\label{frakJ_mu}
    \frak{J}^{(\mu)}_{\pm}=\frac{1+\mu^2}{1-\mu^2} j_{\pm}-\frac{2\mu}{1-\mu^2} j_{\mp} \left(\frac{\mathrm{tr}(j_{\pm}j_{\pm})}{\mathrm{tr}(j_{\mp}j_{\mp})}\right)^{\frac{1}{2}}\,.
\end{equation}
The key observation is that this current has the same structure as the conserved current that one can define for a root-$T\bar{T}$ deformation of the PCM \cite{Borsato:2022tmu}, as indeed the root-$T\bar T$ Lagrangian can be formally obtained from \eqref{L_j_mu} by fixing $\mu=\tanh{\frac{\gamma}{2}}$ (or $y=e^\gamma$) and $H=0$. 
Crucially, this structure allows the current to satisfy the two important conditions
\begin{equation}\label{commutators_mu}
\begin{aligned}
\left[\frak{J}^{(\mu)}_+,\frak{J}^{(\mu)}_-\right]&=\frac{(1+\mu^2)^2}{(1-\mu^2)^2}\left[j_+,j_-\right]\,+\frac{4\mu^2}{(1-\mu^2)^2}\left[j_-,j_+\right]=\left[j_+,j_-\right],
\\
\left[\frak{J}^{(\mu)}_+,j_-\right]-\left[j_+,\frak{J}^{(\mu)}_- \right]&=0\, ,
\end{aligned}
\end{equation}
without setting the auxiliary field $\mu$ on-shell. This is different from the $\nu$-frame and CH cases, where for the latter we still need to use the equation of $\tau$ to satisfy \eqref{LaxPDE}.
Obviously, these relations were already known to be satisfied when $\mu$ is a constant (say $\mu=\tanh{\frac \gamma 2}$ for the root-$T\bar T$ case). The observation in our case is that it does not matter that $\mu$ is not a constant (and actually depends on $j_{\pm}$ through the equations of motion), since the current \eqref{frakJ_mu} is still defined to be conserved. Hence, we can follow the same steps as \eqref{flatness} to conclude that
\begin{equation}\label{Lax_mu}
\frak{L}_{\pm}=\frac{j_{\pm}\pm z\left(\frac{1+\mu^2}{1-\mu^2} j_{\pm}-\frac{2\mu}{1-\mu^2} j_{\mp} \left(\frac{\mathrm{tr}(j_{\pm}j_{\pm})}{\mathrm{tr}(j_{\mp}j_{\mp})}\right)^{\frac{1}{2}}\right)}{1-z^2}\,
\end{equation}
are the components of the Lax connection in the $\mu$-frame. Similarly to what was observed in \cite{Ferko:2024ali} the Lax is a function of both physical and auxiliary fields, which can eventually also be traded for physical fields upon specifying an interaction function $H(\mu)$ in the Lagrangian and exploiting the equation of motion for $\mu$. Having established the connection between all these models we can now write the Lax connection for all of them.
Note that by defining $\mu=\frac{y-1}{y+1}$ in \eqref{Lax_mu} we obtain the expression of the Lax connection for the \emph{y-auxiliary model} which we have argued coincides with the $\mu$-frame in the previous section. Notice finally that we can also write the Lax connection using the Courant-Hilbert function by simply substituting $y=\dot \ell$. This expression must agree with the one in the $\nu$ frame, since we have argued that the Courant-Hilbert model corresponds to a specific choice of $E$ in the $\nu$ frame. Indeed, this can be checked by using the expressions \eqref{v(mu)} on the current \eqref{frac_j_nu_conservation}.  Recalling that $E_\nu=\mu$ we see that we obtain precisely $\frak{J}_\alpha^{(\nu)}\deq\frak{J}_\alpha^{(\mu)}$, therefore showing that the two Lax connections in $\nu$ and $\mu$ frames are the same.

\subsection{Lax in the ($\mu,\rho$)-frame}
In the previous subsection we have argued how to define a Lax connection for the auxiliary field model in the $\mu$-frame, which coincides with the $(\mu,\rho)$-frame model in the limit $\rho=0$. The procedure relies on the commutators \eqref{important_commutators}, which are satisfied as off-shell identities in the $\mu$-frame. This is different from the $\nu$-frame, where \eqref{important_commutators} are only satisfied when the auxiliary fields are on-shell. Additionally, in the $\nu$-frame such commutator identities would be spoiled by a dependency of the interaction function $E$ on the variable $p$, while the extra simplicity of the $\mu$-frame will now allow us to say something more about the integrability of the AFSM even when $E$ depends on both $\nu$ and $p$.

Starting from the combined $(\mu,\rho)$-frame Lagrangian \eqref{L_mu_rho}, one can again construct the conserved current $\frak{J}_\alpha$ in \eqref{current_x12}, which takes the form
\begin{equation}\label{frak_J_mu_rho}
\frak{J}^{(\mu,\rho)}_{\pm}=\frac{1+\mu^2-\rho^2}{(1+\rho)^2-\mu^2} j_{\pm}-\frac{2\mu}{(1+\rho)^2-\mu^2} j_{\mp} \left(\frac{\mathrm{tr}(j_{\pm}j_{\pm})}{\mathrm{tr}(j_{\mp}j_{\mp})}\right)^{\frac{1}{2}}\,.
\end{equation}
The relations \eqref{important_commutators}, needed for the flatness of Lax connections of the form \eqref{Lax_general}, then read
\begin{equation}\label{commutators_mu_rho}
\begin{aligned}
\left[\frak{J}^{(\mu,\rho)}_+,\frak{J}^{(\mu,\rho)}_-\right] &=\frac{(1+\mu^2)^2+\rho^4-2\rho^2(1+\mu^2)-4\mu^2}{((1+\rho)^2-\mu^2)^2}\left[j_+,j_-\right],
\\
\left[\frak{J}^{(\mu,\rho)}_+,j_-\right]-\left[j_+,\frak{J}^{(\mu,\rho)}_- \right]&=0\, ,
\end{aligned}
\end{equation}
where once again we have made no use of any equation of motion. In order to recover the flatness of the Lax \eqref{Lax_general}, given previous discussions, one is led to impose that 
\begin{equation}\label{eq:a=1}
\frac{(1+\mu^2)^2+\rho^4-2\rho^2(1+\mu^2)-4\mu^2}{((1+\rho)^2-\mu^2)^2}=1 \,\, .
\end{equation}
However, the only consistent solution to \eqref{eq:a=1} is $\rho=0$, which recovers the ordinary $\mu$-deformed theory. This is the same result as the one obtained in the $\nu$-frame, where it was shown that $\rho=0$ is a necessary condition to define the Lax connection \eqref{Lax_general}.
A natural question to ask is whether any model with a non-trivial dependency on $\rho$ also admits a zero-curvature representation. To this aim, let us consider a generalised version of \eqref{eq:a=1}, namely
\begin{equation}\label{rho_a}
\frac{(1+\mu^2)^2+\rho^4-2\rho^2(1+\mu^2)-4\mu^2}{((1+\rho)^2-\mu^2)^2}=a^{-2}\,,
\end{equation}
where for $\mu$ and $\rho$ real auxiliary fields, $a$ must necessarily be a real or purely imaginary constant parameter. It is important to notice that we will take \eqref{rho_a} to hold \textit{off-shell}. With this choice, we see that the commutator identity obeyed by $\frak J$ \eqref{commutators_mu_rho} becomes 
\begin{equation} \label{commutator_frakJ}   \left[\frak{J}^{(\mu,\rho)}_+,\frak{J}^{(\mu,\rho)}_-\right]=a^{-2}\left[j_+,j_-\right]\,.
\end{equation}
According to the new condition \eqref{commutator_frakJ}, we can consider the following Lax connection
\begin{equation}\label{Lax_a}
    \frak{L}_{\pm}=\frac{j_{\pm}\pm a z\frak{J}^{(\mu,\rho)}_{\pm}}{1-z^2}
    \,.
\end{equation}
In order to see how flatness of \eqref{Lax_a} is equivalent to the physical fields equations of motion (in analogy with the $a=1$ case), we repeat the steps in \eqref{flatness}:
\begin{align}\label{Mod_flatness}  \mathrm{d}_{\frak{L}}\frak{L}=&~\partial_+\frak{L}_--\partial_-\frak{L}_++\left[\frak{L}_+,\frak{L}_-\right]=
   \notag\\
   =&~\frac{1}{1-z^2}\Bigg(\partial_{+}j_{-}-\partial_{-}j_{+}-az\Bigl(\partial_{+}\frak{J}^{(\mu,\rho)}_{-}+\partial_{-}\frak{J}^{(\mu,\rho)}_{+}\Bigr)
   +\notag\\
   &\qquad \qquad+\frac{\left[j_+,j_-\right]-a^2z^2\left[\frak{J}^{(\mu,\rho)}_+,\frak{J}^{(\mu,\rho)}_-\right]+a z\left[\frak{J}^{(\mu,\rho)}_+,j_-\right]-az\left[j_+,\frak{J}^{(\mu,\rho)}_- \right]}{1-z^2}\Bigg)=
   \notag\\
   =&~\frac{1}{1-z^2}\Bigl(\partial_{+}j_{-}-\partial_{-}j_{+}+[j_+,j_-]-az\left(\partial_{+}\frak{J}^{(\mu,\rho)}_{-}+\partial_{-}\frak{J}^{(\mu,\rho)}_{+}\right)\Bigr)\approx0\,,
\end{align}
 where, as desired, the last line vanishes on-shell by virtue of 
 \begin{equation}\label{frac_j_nu_conservation-2}
    \frak{J}_{\alpha}=-(j_\alpha+2v_{\alpha})\,,
    \qquad\qquad 
     \partial^{\alpha}\frak{J}_{\alpha}\approx 0\,.
\end{equation}
It is worth reiterating that in the final step of \eqref{Mod_flatness} we have used ``$\approx$'' rather than ``$\dapp$'', 
 because flatness of the Lax is equivalent to the physical fields equations of motion \textit{independently} of \eqref{trace_j_relations3}. This will be particularly important in the following discussion.

Notice that the off-shell constraint \eqref{rho_a} is a quadratic equation that we can solve for $\rho(a,\mu)$, de facto eliminating it, and not viewing it as an independent auxiliary field anymore, but rather as a function of $\mu$ and the constant parameter $a$.

It is crucial to notice that imposing \eqref{rho_a} changes the equations of motion \eqref{trace_j_relations3}. Since we are now viewing $\rho = \rho(\mu,a)$, the Lagrangian \eqref{L_mu_rho} becomes a function $\cL = \cL (\mu, \rho(\mu,a))$, and the variation along $\mu$ leads to the new equations of motion 
\begin{equation}
    \del _\mu \cL + \del _\rho \cL \, \del _\mu \rho \,\dot=\, 0\,. 
\end{equation}
The power of working in the $\mu$-frame lies precisely in the fact that imposing \eqref{rho_a} off-shell and thus changing the auxiliary fields equations of motion does not affect the zero-curvature representation of the \textit{physical fields} equations of motion. It is worth stressing again that this is different from the $\nu$-frame, where the current $\frak J$ is only conserved when the auxiliary fields are on-shell. The condition \eqref{rho_a} is equivalent, in the $\nu$-frame, to allowing the interaction function $E$ to satisfy the PDE:
\begin{equation}\label{PDE_E_a}
(1+E_\nu^2)^2+E_p^4-2E_p^2(1+E_\nu^2)-4E_\nu^2=a^{-2}((1+E_p)^2-E_\nu^2)^2\,.
\end{equation}
Therefore, while the standard choice of $a=1$ forces $\rho=E_p=0$, the same condition forces $E$ to obey the differential equation \eqref{PDE_E_a} instead. Notice that when $\mu=E_\nu=0$ the condition on the commutators reduces to imposing $E_p=\frac{a-1}{a+1}$, that is $E=\frac{a-1}{a+1}\mathrm{tr}(v_+v_-)$, which amounts to a (physically irrelevant) rescaling of the PCM Lagrangian after integrating out the auxiliary fields:
\begin{align}\label{L_rescaled}
    \mathcal{L}&=\frac{1}{2} \mathrm{tr}(j_+j_-)+\mathrm{tr}(v_+j_-)+\mathrm{tr}(j_+v_-)+ \mathrm{tr}(v_+v_-)\left(1+\frac{a-1}{a+1}\right)\,,\nonumber\\  &\implies \cL\deq-\frac{1}{2a}\mathrm{tr}(j_+j_-)\,.
\end{align}
Note that in this case, the deformation through $a$ is interpreted as a rescaling of the PCM's Lagrangian, which can be seen as a geometric transformation in the group manifold target space as a rescaling of the Cartan-Killing metric and the target space volume. The reader can, for example, look at \cite{Lacroix:2018njs} for the introduction of such $a$ constant.
For the pure PCM, the presence of a nontrivial parameter $a$ can be interpreted as a marginal, current-current, $J\bar{J}$ deformation. In the case of the auxiliary field sigma model, the presence of an $a\ne 1$ deformation complicates the functional dependence of the model in both the $\mu$ and $\nu$ frames. It would be interesting to further investigate, for generic auxiliary field sigma models, the role of the parameter $a$ as a marginal coupling, for example, by relating it to a marginal deformation driven by an appropriate combination of the $\frak{J}$ current together with $T_{\alpha\beta}$ which would extend the $\mathrm{tr}(j_+j_-)$ operator of \eqref{L_rescaled}. We leave this for future investigations. Let us now move on to the Poisson structure of these models.

\subsection{Poisson structure}
Finding a flat connection is usually not enough to show that a model is classically integrable. The Lax connection is needed to define an infinite tower of commuting conserved quantities and a sufficient condition for the latter requirement has been framed in terms of specific structures that the Poisson brackets of the spatial component of the Lax should satisfy (for details, see the reviews \cite{Torrielli:2016ufi, Driezen:2021cpd}). In various examples, this structure takes the form of a relation involving the Lax itself, an $r$-matrix and a spatial delta-function \cite{Sklyanin:1980ij}
\begin{equation}
    \{\frak{L}_{\sigma,1}(\sigma,z), \frak{L}_{\sigma,2}(\sigma',z')\}= \left[r_{12}(z,z'),\frak{L}_{\sigma,1}(\sigma,z)+\frak{L}_{\sigma,2}(\sigma',z')\right] \delta(\sigma-\sigma').
\end{equation}
In this equation we use the standard notation $X_1=X\otimes \mathds{1}$, $X_2=\mathds{1}\otimes X$ for $X\in \frak{g}$, and $r_{12}\in \frak{g}\otimes \frak{g}$ does not depend on the fields.
However, in the case of the PCM the Poisson brackets have a non-ultralocal form first proposed by Maillet \cite{MAILLET198654}. This means that the previous equation is modified to contain also derivatives of delta-functions. It was already shown in \cite{Ferko:2024ali,Bielli:2024ach} that this is still true when we deform the PCM with auxiliary fields in the $\nu$-frame\footnote{Note that when dealing with constrained Hamiltonian systems one needs to be careful handling the constraints. It is usually necessary to define Dirac brackets instead of the usual Poisson brackets. In our discussion this distinction is irrelevant since we never use the EOM and we refer to \cite{Bielli:2024ach} for more details.}.
The Poisson brackets of the Lax connection in this case have the form \cite{Bielli:2024ach}:
\begin{align}\label{Maillet_brackets}
    \{\frak{L}_{\sigma,1}(\sigma,z),\frak{L}_{\sigma,2}(\sigma',z')\}&= \left[r_{12}(z,z'),\frak{L}_{\sigma,1}(\sigma,z)\right] \delta(\sigma-\sigma')+\nonumber\\
    -&\left[r_{21}(z,z'),\frak{L}_{\sigma,2}(\sigma',z')\right] \delta(\sigma-\sigma')-s_{12}(z,z')\partial_{\sigma}\delta(\sigma-\sigma'),
\end{align}
where we have defined:
\begin{equation} \label{r_matrix}
    r_{12}(z,z')=\frac{C_{12}}{z-z'}\varphi^{-1}(z')\,,\quad \text{with}\quad C_{12}=\gamma^{AB}T_A\otimes T_B \,,\quad \varphi(z)=\frac{z^2-1}{z^2}\,,
\end{equation}
and finally:
\begin{equation}
    s_{12}(z,z')=r_{12}(z,z')+r_{21}(z',z)\,.
\end{equation}
Since this was shown to hold for the $\nu$-frame it is fair to expect that the same calculation should work also for the $\mu$-frame, which describes the same physical models for most choices of the interaction function. This is what we argue in the rest of this section. Interestingly enough this also proves the strong integrability of the other (Courant-Hilbert and $y$-field) auxiliary fields models, which we have already argued are equivalent to the $\mu$-frame of the AFSM.
Let us compute the brackets \eqref{Maillet_brackets} for the Lax connections with a generic parameter $a$. Most intermediate steps, that we omit, can be found in appendix C of \cite{Bielli:2024ach}.
Using the expression for the connection \eqref{Lax_a} we have:
\begin{align}\label{Maillet_j_frakJ}
    \{\frak{L}_{\sigma,1}(\sigma,z),\frak{L}_{\sigma,2}(\sigma',z')\}=&~\frac{1}{(1-z^2)(1-z'^2)}\Bigl(\{j_{\sigma,1}(\sigma),j_{\sigma,2}(\sigma')\}+az\{\frak{J}_{\tau,1}(\sigma),j_{\sigma,2}(\sigma')\}+\nonumber\\& \quad
     +az'\{j_{\sigma,1}(\sigma),\frak{J}_{\tau,2}(\sigma')\}+a^2zz'\{\frak{J}_{\tau,1}(\sigma),\frak{J}_{\tau,2}(\sigma')\}\Bigl)\,.
\end{align}
Hence the three main blocks needed for this equation are the three Poisson brackets involving $j$ and $\frak{J}$. These brackets can be easily obtained from the ones in the $\nu$-frame. The crucial observation to realise this is that the Hamiltonian conjugate momentum in the $\mu$-frame obeys the same relation as in the $\nu$-frame in terms of the $\frak{J}_{\mu}$\footnote{In fact, the relation is different by an overall sign.}. Due to this, the calculation of the brackets on the right-hand side of \eqref{Maillet_j_frakJ} will be the same as those already computed for the $\nu$-frame in \cite{Bielli:2024ach}.
The Hamiltonian conjugate momentum is defined as
\begin{equation}
    \pi_{\mu}=\frac{\partial\mathcal{L}}{\partial \left(\partial_\tau\phi^{\mu}\right)}\,,
\end{equation}
where we are choosing $\phi^{\mu}(\sigma^{\alpha})$, $\mu=1,\dots,\text{dim}(G)\,,\,\,\alpha=1,2\, $, as local coordinates on the group manifold, and $j_{\alpha}=\partial_{\alpha}\phi^{\mu}j_{\mu}{}^A\,T_A$ is the embedding of the field in two dimensions, with $\mathrm{d} \phi^{\mu}=\partial_\alpha\phi^{\mu}\mathrm{d}\sigma^{\alpha}$.
For a generic Lagrangian $\mathcal{L}(x_1,x_2)$, using the notation of \eqref{x_n}, we can compute the momentum as:
\begin{equation}
    \pi_{\mu}=\frac{\partial \mathcal{L}}{\partial x_1}\frac{\partial x_1}{\partial \left(\partial_\tau\phi^{\mu}\right)}+\frac{\partial \mathcal{L}}{\partial x_2}\frac{\partial x_2}{\partial \left(\partial_\tau\phi^{\mu}\right)}=2\frac{\partial \mathcal{L}}{\partial x_1}\partial_\tau\phi^{\nu}G_{\mu\nu}+4\frac{\partial \mathcal{L}}{\partial x_2}\partial_\tau\phi^{\lambda}\partial_\alpha\phi^{\rho}\partial_\beta\phi^{\nu}g^{\alpha\beta}G_{\mu\rho}G_{\nu\lambda}\,,
\end{equation}
where $G_{\mu\nu}=\gamma_{AB}\,j_{\mu}{}^{A}\,j_{\nu}{}^B$. It can be checked again that we can write the time component of the current $\frak{J}_\tau$, defined in \eqref{current_x12}, in terms of the conjugate momentum:
\begin{equation}\label{momentum_frakJ}
    \frak{J}_{\tau}^{B}= \pi_{\mu} j^{\mu}_A \gamma^{AB},
\end{equation}
which is the same relation that was found in the $\nu$-frame \cite{Ferko:2024ali,Bielli:2024ach}, up to a sign. Modulo this sign, the brackets on the right-hand side of \eqref{Maillet_j_frakJ} are the same as in \cite{Bielli:2024ach}:
\begin{align}\label{j_frakJ_Poisson}
    \{\frak{J}_{\tau,1}(\sigma),\frak{J}_{\tau,2}(\sigma')\}=&-\left[\frak{J}_{\tau,2},C_{12}\right]\delta(\sigma-\sigma')\,,\\
     \{\frak{J}_{\tau,1}(\sigma),j_{\sigma,2}(\sigma')\}=&-\left[j_{\sigma,2},C_{12}\right]\delta(\sigma-\sigma')+C_{12}\partial_{\sigma}\delta(\sigma-\sigma')\,,\\
     \{j_{\sigma,1}(\sigma),j_{\sigma,2}(\sigma')\}=&~0\,.
\end{align}
At this point, one needs to use these relations in \eqref{Maillet_j_frakJ} to check if it is possible to define the $r$- and $s$-matrices in \eqref{Maillet_brackets}.
It is not difficult to see that one finds the same objects defined in \cite{Bielli:2024ach, Ferko:2024ali} up to a rescaling of the $r$-matrix (or of the twist function $\varphi$) by $-a$: 
\begin{equation}\label{r-matrix-mu-rho-frame}
    r_{12}(z,z')=-\frac{a\,C_{12}}{z-z'}\varphi^{-1}(z')\,.
\end{equation}
This shows that all the above models, for generic values of $a$, are classically integrable.

\section{Extending the $\mu$-frame to other classes of $2d$ sigma models}\label{s:ext}

One of the remarkable aspects of the $\nu$-frame formulation \cite{Ferko:2024ali} is how amenable it is to being applied to other classes of sigma models. The construction has indeed been extended from the PCM to its non-Abelian T-dual \cite{Bielli:2024khq,Bielli:2025uiv}, to (bi-)Yang-Baxter sigma models \cite{Bielli:2024fnp} and to (semi-)symmetric space sigma models \cite{Bielli:2024oif,Cesaro:2024ipq,Cesaro:2025msv}. It is then natural to imagine that similar $\mu$-frame formulations should exist also for these other classes of models.
In this section we will make this intuition more precise by explicitly constructing $\mu$-frame descriptions of:
\begin{itemize}
    \item AF T-dual models;
    \item AF (bi)-Yang-Baxter deformations;
    \item AF symmetric spaces.
\end{itemize}
We will adopt the partially-unified framework introduced in \cite{Bielli:2025uiv}.

The strategy is identical to the PCM case, namely, we first express the on-shell Lagrangian exclusively in terms of auxiliary fields, $\mu$ and $\rho$, and then show how both the auxiliary and physical fields EOMs arising from this $\mu$-frame Lagrangian are equivalent to the $\nu$-frame ones. Interestingly, we will also see that the extension to the $(\mu,\,\rho)$ frame is not possible for this larger class of sigma models, at least not by treating $\rho$ as an independent auxiliary field.

To mimic the construction presented in previous sections we start from the common $\nu$-frame Lagrangian in \cite{Bielli:2025uiv}
\begin{equation}\label{eq:simplerlag}
\mathcal{L}(\phi,v)=\tfrac{1}{2} \mathrm{tr} \Bigl((\mathcal{K}_{-}+2v_{-})\mathcal{O}_{-}(\mathcal{K}_{+}+2v_{+}) \Bigr)-\mathrm{tr}(v_{-}v_{+})+E(v) \,\, ,
\end{equation}
which, for our purposes, is more conveniently rewritten in the form:
\begin{equation}\label{common-Lagrangian}
\mathcal{L}(\phi,v)=\tfrac{1}{2} \mathrm{tr} (\mathcal{K}_{-}\mathcal{O}_{-}\mathcal{K}_{+})+\mathrm{tr}(v_{-}\mathcal{O}_{-}\mathcal{K}_{+})+\mathrm{tr}(v_{+}\mathcal{O}_{+}\mathcal{K}_{-})+\mathrm{tr}(v_{-}\mathcal{O}_{-}\mathcal{O}_{+}^{-1}v_{+})+E(v) \,\, , 
\end{equation}
where we recall that
\begin{equation}\label{operators-general-form}
\mathcal{O}_{\pm}:=\frac{1}{1\pm M} 
\qquad \text{with} \qquad 
M^{T}=-M 
\quad , \quad \mathcal{O}_{\pm}^{T}=\mathcal{O}_{\mp} \qquad \text{and} \qquad 
\mathcal{O}^{-1}_{\pm}:=1\pm M \, ,
\end{equation}
for $M$ some operator on the underlying Lie algebra, and we refer to section 4.1 in \cite{Bielli:2025uiv} for details on how each of the specific sigma models can be recovered.\footnote{For clarity, we stress that traces containing operators $\cO_\pm$ are traces of two matrices. For example $\mathrm{tr}(v_+ \cO_{\pm} v_-)=(\cO_{\pm})_{B}{}^{C}\,v_+^A v_-^B \mathrm{tr}(T_A T_C)=\mathrm{tr}(v_- \cO_{\mp} v_+)$.}
In analogy with previous sections, looking at the Lagrangian \eqref{common-Lagrangian} it is natural to define the $\nu$ and $p$ variables as
\begin{equation}\label{eq:pnu}
\nu:= \sqrt{\mathrm{tr}(v_{+}^2)\mathrm{tr}(v_{-}^2)} 
\qquad \text{and} \qquad 
p:=\mathrm{tr}(v_{-}\mathcal{O}_{-}\mathcal{O}_{+}^{-1}v_{+})  \,\, ,
\end{equation}
such that $p$ reduces to \eqref{2d-p-nu-definition} for the PCM. The EOM for the auxiliary fields then read
\begin{equation}
\begin{aligned}
\mathcal{O}_{+}\mathcal{K}_{-}+(1+E_{p})\mathcal{O}_{+}\mathcal{O}_{-}^{-1}v_{-}+\nu^{-1}E_{\nu}\mathrm{tr}(v_{-}^2)v_{+} &\deq 0, 
\\
\mathcal{O}_{-}\mathcal{K}_{+}+(1+E_{p})\mathcal{O}_{-}\mathcal{O}_{+}^{-1}v_{+}+\nu^{-1}E_{\nu}\mathrm{tr}(v_{+}^2)v_{-}  & \deq 0,
\end{aligned}
\end{equation}
and using them one can easily compute the on-shell Lagrangian:
\begin{equation}\label{common-on-shell-lagrangian}
\begin{aligned}
\mathcal{L}\deq& +\tfrac{1}{2}p[-1-2E_{p}+E_{p}^2+E_{\nu}^2]+\nu E_{\nu}(E_{p}-1)+E +
\\
& +\tfrac 12 [(1+E_{p})^2-E_{\nu}^2]\mathrm{tr}(v_{-} \mathcal{O}_{-}\mathcal{O}_{+}^{-1}M v_{+}) \,\, .
\end{aligned}
\end{equation}
Unlike for the PCM, we see that \eqref{common-on-shell-lagrangian} does not close into an expression written in terms of the variables $\nu$ and $p$ only. It is therefore natural to introduce an extra ``conjugate'' variable 
\begin{equation}
    q=\mathrm{tr}(v_{-} \mathcal{O}_{-}\mathcal{O}_{+}^{-1}M v_{+}), 
\end{equation}
so as to be able to write \eqref{common-on-shell-lagrangian} as
\begin{equation}
\mathcal{L}\deq \tfrac{1}{2}p[-1-2E_{p}+E_{p}^2+E_{\nu}^2]+\nu E_{\nu}(E_{p}-1)+E +\tfrac 12 q\,[(1+E_{p})^2-E_{\nu}^2]\,.
\end{equation}
Note that we will not assume the interaction function $E$ to depend on $q$, and therefore we will never Legendre transform with respect to this new variable. The reason for this choice will become clear in the following discussion. 
 
The construction of the $(\mu,\,\rho)$-frame Lagrangian for the PCM relied on the fact that two key quantities could be expressed only in terms of derivatives of the interaction function $E$ and the variables $\nu$ and $p$. More specifically, these objects were:
\begin{itemize}
     \item The undeformed Lagrangian $\tr(j_+j_-)$;
     \item $\sqrt {T^{(0)}_{++}T^{(0)}_{--}}\!=\!\sqrt {\mathrm{tr}(j_+^2)\mathrm{tr}(j_-^2)}$, with $T^{(0)}$ the energy momentum tensor of the seed theory.
\end{itemize}
In complete analogy with the PCM, we are now going to compute:
 \begin{itemize}
     \item The building blocks of the undeformed Lagrangian, $\tr (\cK_-\cO^S\cK_+)$ and $\tr (\cK_-\cO^A\cK_+)$, respectively associated to the sigma model kinetic term and $B$-field contribution;
     \item $\sqrt {T^{(0)}_{++}\,T^{(0)}_{--}}$, where $T^{(0)}_{\pm\pm}=\tr (\cK_\pm\cO^S\cK_\pm)$. 
 \end{itemize}
 Note that we are using the notation:
 \begin{equation}
     \cO^S= \frac 12 (\cO _++\cO _-)\, , \quad \cO^A= \frac 12 (\cO_+-\cO_-). 
 \end{equation}
 The calculation is essentially identical to the PCM case, so we will only highlight the key steps. The equations of motion for $v$ arising from \eqref{common-Lagrangian} are 
  \begin{equation}\label{eq:veom}
    \begin{split}
        \cK _++ (1+E_p)\cO_+^{-1}v_++ \tfrac 1\nu E_\nu \tr (v_+^2)\cO _-^{-1}v_-&\deq0,\nl
        \cK _-+ (1+E_p)\cO_-^{-1}v_-+ \tfrac 1\nu E_\nu \tr (v_-^2)\cO _+^{-1}v_+&\deq0 \,\, ,
    \end{split}
\end{equation}
and it is straightforward to check that 
\begin{equation}
    \begin{split}
        \tr (\cK _- \cO _+ \cK _+)
        &\deq (1+E_p)^2 \tr (v_- \cO _+^{-1}v_+)+2\nu E_\nu (1+E_p) + E_\nu ^2 \tr (v_- \cO _+^{-2}\cO _-v_+),\nl 
        \tr (\cK _- \cO _- \cK_+)
        &\deq (1+E_p)^2 \tr (v_- \cO _+^{-2}\cO _-v_+)+2\nu E_\nu (1+E_p) + E_\nu ^2 \tr (v_- \cO _+^{-1}v_+),
    \end{split}
\end{equation}
such that the symmetric and anti-symmetric building blocks of the Lagrangian become
\begin{equation}\label{eq:ckpm}
    \begin{split}
        \tr (\cK _- \cO ^S \cK _+ )&\deq ((1+ E_p )^2 + E_\nu ^2 )p+ 2\nu E_\nu (1+ E_p),\nl 
        \tr (\cK _- \cO^A\cK _+)&\deq (E_\nu ^2 - (1+ E_p)^2 )q.
    \end{split}
\end{equation}
Similarly, the undeformed stress tensor components reduce to 
\begin{equation}\label{eq:T0}
    \begin{split}
        T^{(0)}_{\pm\pm}=\tr (\cK_\pm\cO ^S\cK _\pm) &\deq \tr (v_\pm^2 )\left ( (1+ E_p)^2 + E_\nu ^2 +2 \nu ^{-1}(1+ E_p)E_\nu p\right ),
    \end{split}
\end{equation}
so that 
\begin{equation}
    \sqrt {T^{(0)}_{++}T^{(0)}_{--}}\deq \nu ((1+E_p)^2+ E_\nu ^2)+ 2 p E_\nu (1+E_p).
\end{equation}
We see that going on-shell for $v$ all these objects can be expressed exclusively in terms of $E_{\nu},E_{p}$ and $\nu,p$. Before proceeding with the Legendre transformation, it is worth briefly commenting about the possibility of choosing $E_q\neq 0$. Had we done this, the equations of motion \eqref{eq:veom} would have picked up terms of the form 
\begin{equation}
    \ldots \pm E_q\cO_\pm ^{-1}Mv_\pm\ldots.
\end{equation}
One can then check that the expressions for $T^{(0)}_{\pm\pm}$ also get modified, in particular, terms proportional to $E_q^2$ are given by
\begin{equation}
    \tr(\cK_\pm\cO^S\cK_\pm)\deq\ldots - E_q ^2 \tr (v_\pm M^2 v_\pm).
\end{equation}
Hence, one can see that the quantity $\sqrt {T^{(0)}_{++}T^{(0)}_{--}}$ would in general not close into an expression written in terms of derivatives of $E$ and $\nu,p,q$ only. It would be interesting to understand whether ``closure'' could be achieved by introducing a tower of extra variables 
\begin{equation}
    \nu_{k,l}=\tr(v_+M^kv_+)\tr(v_-M^lv_-) \,\, ,
\end{equation}
which is however beyond the scope of this work and is left as a possible future investigation. 

Following the construction of section \ref{s:murhofr} we now perform the Legendre transform \eqref{Legendre_mu_mutilde}:
\begin{equation}
    H = E- \nu E_\nu - p E_p, \qquad \mu = E_\nu, \qquad \rho = E_p, 
\end{equation}
which can be substituted into \eqref{eq:ckpm} and \eqref{eq:T0} to give 
 \begin{equation}\label{eq:legtraces}
     \begin{split}
         \tr (\cK _- \cO ^S\cK _+)&\deq - 2 \mu H_\mu - H_\rho ( \mu ^2 + (1+\rho) ^2 ),\nl 
         \tr (\cK _- \cO ^A\cK _+)&\deq q (\mu ^2 -(1+\rho)^2 ),\nl 
         \sqrt {T^{(0)}_{++}T^{(0)}_{--}}&\deq -2 H _\rho \mu (1+ \rho )- H_\mu (\mu ^2 +(1+ \rho) ^2 ). 
     \end{split}
 \end{equation}
The on-shell Lagrangian then becomes:
  \begin{equation}\label{extended_mu_L}
      \cL = \frac 1{(1\!+\!\rho)^2\!-\!\mu ^2}\left(\! \mu \sqrt {T^{0}_{++}T^{(0)}_{--}}+\frac 12 (\rho^2 \!-\!\mu^2 \!-\!1)\tr (\cK_+\cO ^S\cK_-)\!\right) + \frac 12 \tr (\cK _+\cO^A\cK_-)+ H.
\end{equation}
Crucially, it is easy to check that variations of \eqref{extended_mu_L} along $\mu$ and $\rho$ impose the EOM
  \begin{equation}
      \begin{split} 
          H_\mu &\deq\frac {2\mu (1+ \rho ) \tr (\cK _+ \cO ^S\cK _-)- ( \mu ^2 +(1+ \rho )^2 )\sqrt {T^{(0)}_{++}T^{(0)}_{--}}}{(\mu^2-(1+\rho)^2)^2}, \nl 
         H _\rho &\deq \frac {2\mu (1+ \rho ) \sqrt {T^{(0)}_{++}T^{(0)}_{--}}- ( \mu ^2 + (1+\rho) ^2 )\tr (\cK _+ \cO ^S\cK _-)}{(\mu^2 -(1+\rho)^2)^2}, 
      \end{split}
  \end{equation}
 which coincide with the on shell  expressions for $H_{\mu}$ and $H_\rho$ derived by rearranging \eqref{eq:legtraces} (which is why we have used the sign ``$=$'' in \eqref{extended_mu_L}).
 
We have thus obtained the combined $(\mu,\rho)$-frame Lagrangian describing in a unified way the deformations of all the extensions of the PCM mentioned at the beginning of this section. Firstly, it should be noted that, as expected, the antisymmetric building block $\cO^A$ of the undeformed Lagrangian, corresponding to a $B$-field contribution, is left completely untouched by the change of frame. In the Lagrangian \eqref{extended_mu_L} we see this by the fact that $\mathrm{tr}(\cK_+\cO^A\cK_-)$ is not coupled to either auxiliary field. Also, it is clear that, due to the definition of $p$, admitting that $E$ depends on $p$ for the T-dual and bi-YB cases means allowing the interaction function to depend on the physical fields through $\cO_\pm$ inside $p$. Equivalently, after the Legendre transformation this would imply that $\rho$ depends on the physical fields. Although this fact does not seem to play a role when showing that the $(\mu,\,\rho)$-frame reproduces the correct auxiliary field equations of motion, it is crucial in our analysis of the physical fields EOMs. More specifically, we show in appendix \ref{a:physeom} that if we formally vary the physical fields in \eqref{extended_mu_L} only through their \textit{explicit} appearance in $\sqrt {T^{(0)}_{++}T^{(0)}_{--}}$, $\tr (\cK _+\cO^A\cK_-)$ and $\tr (\cK_+\cO ^S\cK_-)$, the resulting equations of motion are equivalent to the $\nu$-frame ones if and only if $\rho =0$ (note that this is only true for T-dual models and (bi)-Yang-Baxter, while the symmetric space extension to the $(\mu,\,\rho)$-frame is identical to the PCM because the operator $M$ vanishes, i.e. $\mathcal{O}_{\pm}=1$ and there is no $B$-field, such that $p$ is independent of the physical fields). It would be interesting to understand whether there is a consistent way to vary $\rho$ with respect to the physical fields. Of course, in this case, the idea of treating $\rho$ as an auxiliary field after the Legendre transform loses its meaning. It could nonetheless give rise to new examples of integrable models where the interaction function also depends on the physical fields. Perhaps this could be achieved by first varying $p$ in \eqref{eq:pnu}, and then re-expressing the variation in terms of the physical fields, $\mu$, $p$ and $E$ as we do for conserved and flat currents which typically appear in these models. One could then attempt to ``prescribe'' this variation as a starting point in \eqref{extended_mu_L} and compute the associated equations of motion. We will not consider this here, instead, we simply show in \ref{a:physeom} that treating $\rho$ as an auxiliary is (as expected) inconsistent. 
 
 Before moving on to the conclusions, we will argue why the integrability of these models is immediate. Both in the $\nu$ and $\mu$-frames, weak integrability (namely the construction of a Lax connection) only relies on \textit{two facts}:
 \begin{enumerate}
     \item The EOM are equivalent to the flatness of a current and the conservation of another. 
     \item Let us denote the flat current by $\cA$ and the conserved one by $\cB$, following the notation of \cite{Bielli:2025uiv}. In order to construct the Lax, these currents must satisfy a set of \textit{algebraic} conditions, namely
     \begin{equation}\label{eq:intprops}
         \begin{split}
            \left[\cB_+,\cB_-\right] &=a^{-2}\left[\cA_+,\cA_-\right],\nl
            \left[\cB_+,\cA_-\right] &=\left[\cA_+,\cB_- \right], 
         \end{split}
     \end{equation}
     where $a$ is a constant. 
 \end{enumerate}
Let us temporarily take $a=1$: given these two simple properties, it was shown in \cite{Bielli:2024ach,Bielli:2024fnp,Bielli:2024khq,Bielli:2024oif} that the Lax connection can be written in a canonical way for all these AFSMs in terms of these currents. However, as we have seen here and in appendix \ref{a:physeom}, the currents we find in the $\mu$-frame are \textit{exactly} the same as those in the $\nu$-frame when going on-shell for the auxiliary fields, hence the properties \eqref{eq:intprops} hold. Thus, the form of the Lax connections for the (bi)-Yang-Baxter and T-dual deformed models will be identical to the expressions written in \cite{Bielli:2024ach,Bielli:2024fnp,Bielli:2024khq,Bielli:2024oif} where the currents are simply re-written in terms of $\mu$-frame objects, and no ``bare'' $v$'s appear. We then conclude that all these models will be integrable with exactly the same Lax connections as in the $\nu$-frame, in complete analogy with the PCM. The case $a\neq1$ extends trivially to symmetric space models by following the $(\mu,\,\rho)$-frame analysis of the PCM presented in section \ref{s:murhofr}. We recall the universal form of the Lax connections for these models:
\begin{equation}\label{eq:univLax}
    \mathfrak L _\pm= l^\pm _1(z)\frac {\cA_\pm\pm za\cB_\pm }{1+z}+l^\pm_2(z)\Theta (v_\pm)+l_3^\pm(z)\cC_\pm,
\end{equation}
where $\Theta:\fg\rightarrow \fg$ and  $l^\pm_{1,2,3}:\mathbb C \rightarrow\mathbb C$ are model dependent functions, while the current $\cC$ defines a covariant derivative 
\begin{equation}
    D_\pm= \del _\pm + [\cC_\pm,-],
\end{equation}
where relevant. As explained, we do not need to invoke the specific functional form of these objects as they are identical to the $\nu$-frame quantities described in equations (4.16$-$4.19) of \cite{Bielli:2025uiv}. 
 
\section{Conclusions and Outlook}
Models with auxiliary fields have been intensively studied recently, among other things, in the context of extensions of electrodynamics. They have proven to be a useful tool for understanding duality-invariance of these theories \cite{Ferko:2023ruw,Ferko:2023wyi, Russo:2024llm}, finding conditions for their causality \cite{Russo:2024llm} (see also \cite{Babaei-Aghbolagh:2026vkm}) and relations to energy conditions \cite{Russo:2024xnh}. They have also played an important role in the study of stress-tensor deformations of electrodynamics \cite{Ferko:2023ruw,Ferko:2023wyi, Babaei-Aghbolagh:2024uqp,Babaei-Aghbolagh:2025cni}.
In section \ref{s:4d} we have focused on the structure of the Ivanov-Zupnik auxiliary field model of electrodynamics \cite{Ivanov:2001ec,Ivanov:2002ab}. Recent works concerning similar models \cite{Russo:2025fuc,Kuzenko:2025gvn} suggested a relation with the formalism of this model. We have described precisely this correspondence connecting the $\nu$-frame and the $\mu$-frame with the Courant-Hilbert solution of \eqref{PDE} --- as already anticipated in appendix A of \cite{Ferko:2023wyi} --- and the related scalar field auxiliary model of \cite{Russo:2025fuc}. Finding this relation is also important to connect several results that have been understood separately in these different formalisms. In sections \ref{2D_muframe}, \ref{s:int} and \ref{s:ext} we focused on $2d$ models. More specifically, we showed that in the case of the principal chiral model, the auxiliary field sigma model in the $\nu$-frame, constructed by coupling the PCM to a vector auxiliary field $v$, is equivalent, upon performing a Legendre transform, to its $\mu$-frame formulation which instead couples the theory to a scalar auxiliary field. This way we have also shown the relation between the $\nu$-frame AFSM and the CH/$y$-field description of integrable sigma models recently discussed in \cite{Fukushima:2025tlj, Babaei-Aghbolagh:2025uoz, Babaei-Aghbolagh:2025hlm}. In section \ref{s:int} we discussed the weak and strong integrability of these models. In addition, we found that there exists an entire additional family of integrable deformations where an extra auxiliary field $\rho$ is introduced, and the Lax connection is modified by the introduction of an extra constant parameter $a$. These theories correspond to allowing a particular dependence on $\tr(v_+v_-)$ in the $\nu$-frame theory. Interestingly, in this case, we find that the flatness of the Lax connection reduces to a particular non-linear algebraic equation relating $\mu$ and $\rho$, which translates to a non-linear PDE in the variables $\nu$ and $p$ which the interaction function $E$ has to satisfy. The $(\mu,\, \rho)$-frame formulation is especially convenient to describe the integrable models where $E_p\neq 0$, since the algebraic relation involving $\mu $ and $\rho $ which ensures integrability corresponds to a non-linear PDE in the $\nu$-frame. Obviously, the PDE is a significantly more complicated system to solve. After constructing the $(\mu,\,\rho )$-frame for the PCM we repeated the analogous construction for several more AFSMs via the unified framework introduced in  \cite{Bielli:2025uiv}. We found that in this case, the conjugate variable $p$ picks up a non-trivial dependence on the physical fields via the operators $\cO _\pm$. Due to this extra dependence, we see that while the $\mu$-frame extends straightforwardly to all these models, the $(\mu,\,\rho)$-frame does not, at least not by treating the variable $\rho$ as an auxiliary field. 

We believe that the above findings open several potential future directions to explore for deformations in various dimensions. Here we focus on those related to $2d$.
\begin{itemize}
    \item The $(\mu,\,\rho)$-frame models described in this paper consist of an entire new family of integrable $2d$ QFTs. The power of the auxiliary field formalism lies precisely in the fact that all the ``integrability data'' are relegated to the \textit{structure} of the interaction function and the couplings of physical and auxiliary fields. While this is particularly powerful, it is also complicated to realise concrete examples of interesting models, namely to explicitly construct examples of interaction functions. For instance, thanks to Zamolodchikov's point splitting argument \cite{Zamolodchikov:2004ce,Smirnov:2016lqw} we know that any deformation triggered by a flow which is quadratic in some conserved current will be well defined quantum mechanically. The same cannot be said for some arbitrary function of conserved currents. In this sense, the set of quantum mechanically interesting deformations is (as expected) smaller than the naive enormous family of flows which can be constructed classically. The analysis in \cite{Bielli:2025uiv} showed that even the explicit construction of interaction functions encoding Smirnov-Zamolodchikov flows is rather laborious, and it could only be achieved perturbatively in the irrelevant coupling. It would be interesting to construct an explicit ``sensible'' example of $(\mu,\, \rho)$-frame deformed model, integrate out the scalar auxiliary fields and try to understand what kind of deformation it encodes and whether they can be useful to study field theories at the quantum level. This is particularly interesting since such deformations mix irrelevant and (classically)marginal operators. 
    
    \item The entire $\mu$-frame construction relies on the fact that the interaction function $E$ in the $\nu$-frame only depends on $\tr (v^2)$ and not higher traces. This translates to the flows being triggered by functions of the energy-momentum tensor. Hence, these are often referred to as $T\ov T$-like deformations. Accordingly, the $(\mu, \, \rho)$-frame is a further extension of these. 
    A natural question is to ask whether $\mu$-frame realisations exist also for higher-spin deformations, Smirnov-Zamolodchikov-type deformations, engineered in the $\nu$-frame. This might happen by taking Legendre transforms with respect to a set of Lorentz invariant variables constructed out of products of $\tr(v_+^n)$ and $\tr(v_-^n)$. Analogously to the new role of integrable AFSM discovered in this work for non-trivial dependencies upon $\tr(v_+v_-)$, it would be interesting to see whether larger classes of integrable field theories could extend the sufficient conditions for integrability employed in \cite{Bielli:2024ach,Bielli:2025uiv}. To this end, it might be intriguing to see whether higher-spin analogues and extensions of the $(\mu,\rho)$-frame exist.

    \item As mentioned at the end of section \ref{s:ext}, when attempting to extend the $(\mu,\, \rho )$-frame construction to all AFSMs one finds that the conjugate variable $p$ depends on both auxiliary and physical fields. This implies that one cannot just treat $\rho$ as an auxiliary field in the $(\mu,\,\rho)$-frame. In section \ref{s:ext} and appendix \ref{a:physeom} we simply note that the naive treatment is inconsistent and set $\rho \!=\!0$. It seems natural to ask whether one can nonetheless consistently deform using $\rho$. In principle, one could vary $p\! =\! \tr (v_- \cO_-\cO_+^{-1}v_+)$ and attempt to express the variation in terms of $E$, $\nu$, $p$, and physical fields as we have done for the conserved/flat currents. One could then postulate this variation in the $\mu$-frame and attempt to incorporate $\rho$-deformations not as independent auxiliary fields, but rather as objects depending on the physical fields. 
  
\end{itemize}
We leave all these interesting investigations to future work.

\section*{Acknowledgements}

We are grateful to Christian Ferko for discussions and collaboration related to this project.
G.\,T.-M., and M.\,G. have been supported by the Australian Research Council (ARC) Discovery Project DP240101409, and a faculty start-up funding of UQ’s School of Mathematics and Physics. 
N.\,B. is supported by a postgraduate scholarship at the University of Queensland. 
D.B. has been supported by a Young Scientist Training (YST) Fellowship from Asia Pacific Center for Theoretical Physics (APCTP) and the Thailand NSRF via PMU-B, grant number B13F680083.

\appendix
\addtocontents{toc}{\protect\setcounter{tocdepth}{1}}

\section{Equality of the EOM in different frames}\label{app_a}
Going from the $\nu$-frame to the $\mu$-frame Lagrangian requires using the equations of motion (EOM) for the auxiliary fields and the on-shell expression of the Lagrangian itself. It is not obvious that the equations of motion for the physical fields stay the same in the two frames. Clearly, if they were not the same, we would be looking at two completely different systems, and the idea of a change of frame would not make sense. Let us explicitly show how to check that these equations of motion are the same for the two-dimensional case.

In the $\nu$-frame the equations of motion for $j_\alpha$ and $v_\alpha$ are respectively \cite{Bielli:2024ach}:
\begin{equation}\label{general_eom_nu}
\begin{aligned}
    \partial_+\left(j_-+2v_- \right)+\partial_-\left(j_++2v_+ \right)&\approx2(\left[v_+,j_-\right]+\left[v_-,j_+\right])\,,\\
    j_++v_+(1+E_p)+v_-\frac{\sqrt{\nu_{+2}}}{\sqrt{\nu_{-2}}}E_\nu&\deq0\,,\\
    j_-+v_-(1+E_p)+v_+\frac{\sqrt{\nu_{-2}}}{\sqrt{\nu_{+2}}}E_\nu&\deq0\,.
\end{aligned}
\end{equation}
Let us show that the EOM are the same for the combined $(\mu,\rho)$-frame model.
The equation for the physical field $\mathrm{g}$ can be found from the Lagrangian \eqref{L_mu_rho} 
\begin{align}\label{eom_mu}
    \partial_+\Bigl(-\frac{1+\mu^2-\rho^2}{(1+\rho)^2-\mu^2}j_- +\frac{2\mu}{(1+\rho)^2-\mu^2}\frac{\mathrm{tr}(j_-^2)^{\frac{1}{2}}}{\mathrm{tr}(j_+^2){\frac{1}{2}}}j_+\Bigl)&+ \nonumber\\
    \partial_-\Bigl(-\frac{1+\mu^2-\rho^2}{(1+\rho)^2-\mu^2}j_+&+\frac{2\mu}{(1+\rho)^2-\mu^2}\frac{\mathrm{tr}(j_+^2)^{\frac{1}{2}}}{\mathrm{tr}(j_-^2){\frac{1}{2}}}j_-\Bigl)\approx0\,,
\end{align}
where we used the notation $\mathrm{tr}(v_{\pm}^2)=\nu_{\pm2}$.
In the $\nu$-frame one can then manipulate the second and third equations in \eqref{general_eom_nu} to rewrite $v_{\pm}$ only in terms of $j_{\pm}$ and $E_\nu=\mu$. Start by substituting $v_-$ into the equation for $v_+$:
\begin{align}
    v_+(1+E_p)\deq&-j_+-v_-\frac{\sqrt{\nu_{+2}}}{\sqrt{\nu_{-2}}}E_\nu= -j_+-(1+E_p)^{-1}\left(-j_--v_+\frac{\sqrt{\nu_{-2}}}{\sqrt{\nu_{+2}}}E_\nu\right)\frac{\sqrt{\nu_{+2}}}{\sqrt{\nu_{-2}}}E_\nu\deq
    \nonumber\\
    \deq&-j_++\frac{j_-\frac{\sqrt{\nu_{+2}}}{\sqrt{\nu_{-2}}}E_\nu}{1+E_p}+v_+\frac{E_\nu^2}{1+E_p}\,.
\end{align}
Since one can do the same thing for $v_-$, the following on-shell expressions are obtained:
\begin{equation}\label{v_pm}
    v_{\pm}\deq\frac{-j_{\pm}(1+E_p)+j_{\mp}\nu_{\pm2}^{\frac{1}{2}}\nu_{\mp2}^{-\frac{1}{2}}E_\nu}{(1+E_p)^2-E_\nu^2}\,.
\end{equation}
Squaring the latter and taking the trace one then obtains
\begin{equation}
\nu_{\pm2}\left((1+E_p)^2-E_\nu^2\right)^2\deq\mathrm{tr}(j_{\pm}^2)(1+E_p)^2-2\mathrm{tr}(j_+j_-)\nu E_\nu (1+E_p)\nu_{\mp2}^{-1}+\mathrm{tr}(j_{\mp}^2)\nu^2 E_\nu^2 \nu_{\mp2}^{-2}\,.
\end{equation}
Multiplying the two sides of this relation by $\nu_{\mp2}$ one can see that the left-hand side of both equations is the same. One can therefore equate the two right-hand sides, leading to an equation without the $\mathrm{tr}(j_+j_-)$ term:
\begin{equation}
    \mathrm{tr}(j_+^2)\nu_{-2} (1+E_p)^2+\mathrm{tr}(j_-^2)\nu^2 E_\nu^2 \nu_{-2}^{-1} \deq\mathrm{tr}(j_-^2)\nu_{+2}(1+E_p)^2+\mathrm{tr}(j_+^2)\nu^2 E_\nu^2 \nu_{+2}^{-1}\,,
\end{equation}
which eventually suggests that
\begin{equation}
    \frac{\mathrm{tr}(j_+^2)}{\nu_{+2}}\left(\nu^2(1+E_p)^2-\nu^2 E_\nu^2\right)\deq\frac{\mathrm{tr}(j_-^2)}{\nu_{-2}}\left(\nu^2(1+E_p)^2-\nu^2 E_\nu^2\right) \, \implies \frac{\mathrm{tr}(j_+^2)}{\mathrm{tr}(j_-^2)}\deq \frac{\nu_{+2}}{\nu_{-2}}\,.
\end{equation}
Finally, one can rewrite \eqref{v_pm} in terms of $j_\pm$ and $E_\nu$ only as
\begin{equation}\label{v(mu)}
    v_{\pm}\deq\frac{-j_{\pm} (1+E_p)+j_{\mp}\mathrm{tr}(j_{\pm}^2)^{\frac{1}{2}}\mathrm{tr}(j_{\mp}^2)^{-\frac{1}{2}}E_\nu}{(1+E_p)^2-E_\nu^2}\,.
\end{equation}
Substituting these expressions inside the first equation in \eqref{general_eom_nu} it is easy to find agreement with \eqref{eom_mu} provided that $\mu=E_\nu$ and $\rho=E_p$.
From this calculation, one can conclude that the two frames describe the same physical system up to the Legendre transformation sending $(E,\nu,p)$ into $(H,\mu, \rho)$. The same logic holds for the T-dual and (bi-)YB sigma models, whose equations of motion in the $\mu$-frame are analysed in detail in appendix \ref{app_C}.

\section{Stress-energy tensor in the $\mu$-frame}\label{app_b}
The transformation from the $\nu$ to the $\mu$ frame connects two different Lagrangians, which describe exactly the same physical system. One would hence expect that, when the auxiliary fields are on-shell, it should be possible to obtain the stress-energy tensor in one frame by simply applying the frame transformation rules to the stress-energy tensor in the other frame. In this paper, we only consider invariant quantities, built out of the stress-energy tensor, which were already considered in the $\nu$-frame \cite{Ferko:2024ali}\footnote{Notice that the quantity $\nu$ in this reference is here denoted by $\nu^2$.}:
\begin{equation}\label{stress_energy_nu}
T^{\alpha}{}_{\alpha}\deq2(E-\nu E_{\nu})\,, 
\qquad \qquad T^{\alpha\beta}T_{\alpha\beta}\deq\frac{1}{2}\nu^2\left(1-E_\nu^2\right)^2+2(E-\nu E_\nu )^2\,.
\end{equation}
Let us compute these two quantities explicitly and show how they can be obtained by Legendre transforming \eqref{stress_energy_nu}. The stress-energy tensor for $2d$ sigma models reads:
\begin{equation}
T_{\alpha\beta}=g_{\alpha\beta}\mathcal{L}-2\frac{\partial \mathcal{L}}{\partial g^{\alpha\beta}}.
\end{equation}
It is convenient to follow the same notation as in \cite{Borsato:2022tmu}, introducing the quantities
\begin{equation} \label{x_n}
(x_1)_{\alpha\beta}=\mathrm{tr}(j_\alpha j_\beta)\,,
    \qquad\qquad (x_n)_{\alpha\beta}=(x_1)_{\alpha\gamma_1}(x_1)^{\gamma_1}_{\,\,\,\gamma_2}\dots(x_1)^{\gamma_n}_{\,\,\,\beta},
\end{equation}
as one can build the action in the $\mu$-frame from
\begin{equation}
(x_n)^{\alpha}_{\,\,\,\alpha}=x_n\,,
\qquad\quad
x_1=-\mathrm{tr}(j_+j_-)\,,
\qquad\quad
2x_2-x_1^2=\mathrm{tr}(j_-^2)\mathrm{tr}(j_+^2)\,.
\end{equation}
Notice finally that one also has the following relations
\begin{equation}
\frac{\partial x_1}{\partial g^{\alpha\beta}}=(x_1)_{\alpha\beta}\,,
\qquad\qquad
\frac{\partial x_2}{\partial g^{\alpha\beta}}=2(x_2)_{\alpha\beta}\,\,.
\end{equation}
The expression of the stress-energy tensor is then:
\begin{equation}
T_{\alpha\beta}\!=\!\!\left(\!\!f(\mu)x_1\!+\!g(\mu)\sqrt{2x_2\!-\!x_1^2}\!+\!H(\mu)\!\!\right)g_{\alpha\beta}-2f(\mu)(x_1)_{\alpha\beta}-2g(\mu)\frac{2(x_2)_{\alpha\beta}\!-\!x_1(x_1)_{\alpha\beta}}{\sqrt{2x_2-x_1^2}}\,,
\end{equation}
where we defined $f(\mu)=\frac{1}{2}\frac{1+\mu^2}{1-\mu^2}$ and $g(\mu)=\frac{\mu}{1-\mu^2}$, constant under variations of the metric.

The first of the two invariants in \eqref{stress_energy_nu} is easy to check:
\begin{equation}
    T^\alpha_{\,\,\,\alpha}= T_{\alpha\beta}g^{\alpha\beta}=2\left(fx_1+g\sqrt{2x_2-x_1^2}+H\right)-2fx_1-2g\sqrt{2x_2-x_1^2}=2H\,,
\end{equation}
which is exactly the Legendre transform of $E$.

The second invariant requires a slightly longer calculation. The square of $T_{\alpha\beta}$ reads
\bea
T_{\alpha\beta}T^{\alpha\beta}&=&
+4\Bigg(f^2 x_2+g^2\frac{4x_4+x_1^2x_2-4x_1x_3}{2x_2-x_1^2}
+2fg \frac{2x_3-x_1x_2}{\sqrt{2x_2-x_1^2}}\Bigg)
\non\\
&&
-2\Bigg(fx_1+g\sqrt{2x_2-x_1^2}\Bigg)^2
+2H^2 \, ,
\eea
and one needs to express the invariants made out of three and four traces in terms of $x_1$ and $x_2$. By the Cayley-Hamilton theorem, the matrix $(x_1)_{\alpha\beta}$ solves its own characteristic equation. To simplify the notation let us denote $(x_n)_{\alpha\beta}=\hat x_n$, such that one has
\begin{equation}
    \hat x_1^2-\mathrm{tr}(\hat x_1)\hat x_1+\frac{1}{2}\left(\mathrm{tr}(\hat x_1)^2-\mathrm{tr}\left(\hat x_1^2\right)\right)\mathbb{I}_{2\times2}=0\,.
\end{equation}
To find an equation for $x_3$ one can multiply this equation by $(x_1)_{\alpha\beta}$ and take the trace, while for $x_4$ one can simply square the equation:
\begin{equation}
\begin{aligned}
x_3&\!=\!\mathrm{tr}(\hat x_1 \hat x_1^2)\!=\! \mathrm{tr}\!\left(\!\hat x_1(\mathrm{tr}(\hat x_1)\hat x_1-\frac{1}{2}\left(\mathrm{tr}(\hat x_1)^2-\mathrm{tr}\left(\hat x_1^2\right)\right)\mathbb{I}_{2\times2})\!\right)\!=\frac{3}{2}x_1 x_2-\frac{1}{2}x_1^3\,,
\\
x_4&\!=\!\mathrm{tr}(\hat x_1^2 \hat x_1^2)\!=\! \mathrm{tr}\!\left(\!\!\left(\mathrm{tr}(\hat x_1)\hat x_1-\frac{1}{2}\left(\mathrm{tr}(\hat x_1)^2-\mathrm{tr}\left(\hat x_1^2\right)\right)\mathbb{I}_{2\times2}\right)^2\right)\!=\frac{1}{2}x_2^2+x_1^2x_2-\frac{1}{2}x_1^4\,.
\end{aligned}
\end{equation}
With these formulae, it is possible to rewrite the square of the stress-tensor as
\begin{align}
&T_{\alpha\beta}T^{\alpha\beta}\!=\!-2\left(fx_1+g\sqrt{2x_2-x_1^2}\right)^2\!+\!2H^2 \!+\!4\left(f^2 x_2+g^2x_2+2fg x_1\sqrt{2x_2-x_1^2}\right)\,.
\end{align}
Substituting now the expressions for $f$ and $g$ one arrives at
\begin{equation}
T_{\alpha\beta}T^{\alpha\beta}=-\frac{1}{2}x_1^2+\frac{(1+\mu^2)^2}{(1-\mu^2)^2}x_2+\frac{2\mu(1+\mu^2)x_1\sqrt{2x_2-x_1^2}}{(1-\mu^2)^2}+2H^2\,,
\end{equation}
which upon exploiting the EOM for the field $\mu$ in \eqref{trace_j_relations2} finally becomes
\begin{equation}\label{stress_energy_mu}
T_{\alpha\beta}T^{\alpha\beta}\deq\frac{1}{2}(H_\mu)^2(1-\mu^2)^2+2H^2\,.
\end{equation}
Once again, it can be checked that this is exactly the same as transforming the invariant in \eqref{stress_energy_nu} by using the maps between the $\nu$ and $\mu$ frames.

\section{EOM for the T-dual and (bi)-YB sigma models}\label{app_C}
Here we collect some calculations related to section \ref{s:ext}.
The idea is that the $\mu$-frame Lagrangian we have found should describe the same system as the Lagrangian we started from \eqref{eq:simplerlag}. The way to show this is to check that the equations of motion of the physical fields match in both frames. This has already been done for the combined $(\mu,\rho)$-frame for the AF-PCM in appendix \ref{app_a}, which is also enough to be convinced that the same EOM are found in the case of AF symmetric space sigma models. The two less trivial cases are the T-dual and (bi-)Yang-Baxter deformed auxiliary field models, whose EOM have a more complicated structure. Notice also that for these two cases we fix $\rho=0$, as a $\rho$ dependence would imply that the interaction function $E$ depends on the physical fields through $\cO_\pm$. In the following, we report the calculations showing that these equations match.

\subsection{AF T-dual PCM}\label{a:physeom}

\paragraph{$\nu$-frame.} Let us show what are the equations of motion computed from \eqref{eq:simplerlag}. Recall that for the T-dual model the identifications are $\cK_{\pm}=\pm\partial_{\pm}X$ and $\cO_{\pm}=\frac{1}{1\!\pm\!\mathrm{ad}_{X} }$. In the following we ignore a possible dependence on $p$ for the interaction function ($\rho=0$). The equations of motion for the physical fields can be shown to be:
\begin{align}\label{Tdual_eom_nu}
    \partial_-\Bigl(\cO_-\left(\partial_+ X +2v_+\right)\Bigl)-&\partial_+\Bigl(\cO_+\left(-\partial_- X +2v_+\right)\Bigl)+\nonumber\\&+\left[\cO_+\left(-\partial_- X +2v_+\right),\cO_-\left(\partial_+ X +2v_+\right)\right]=0\,.
\end{align}
These are easy to compute using that the variation of $\cO_\pm$ with respect to $X$ take the form
\begin{equation}
    \delta \cO_\pm(\,\cdot\,)=\mp\cO_\pm\delta \,\mathrm{ad}_X( \cO_\pm\,\cdot\,)=\mp\cO_\pm [\delta X, \cO_\pm\,\cdot\,]\,.
\end{equation}
Equation \eqref{Tdual_eom_nu} imposes the flatness equation for the current $\tilde{\frak{J}}_\pm$, thus defined as:
\begin{equation}
\tilde{\frak{J}}_\pm=\pm\cO_\mp\left(\partial_\pm X \pm2v_\pm\right)\,.
\end{equation}
Similarly to what was shown in appendix \ref{app_a} for the AF-PCM, one can here use the EOM of the auxiliary fields to express $v_\pm$ as:
\begin{equation}\label{extended_v(j,mu)}
    v_\pm=\frac{-\cO_{\pm}\left(\cK_\pm -\cK_\mp\mu \sqrt{\frac{T^{(0)}_{\pm\pm}}{T^{(0)}_{\mp\mp}}}\right)}{1-\mu^2}=\frac{\mp\cO_{\pm}\left(\partial_\pm X+\mu \,\partial_\mp X \sqrt{\frac{T^{(0)}_{\pm\pm}}{T^{(0)}_{\mp\mp}}}\right)}{1-\mu^2}\,,
\end{equation}
which tells us that the physical field equations of motion in the $\nu$-frame can be rewritten as the flatness of the current written only in terms of $\mu=E_\nu$ and $X$
\begin{equation}\label{frakJ_Tdual_nu}
\tilde{\frak{J}}_\pm=\mp\left( \frac{\mu^2}{1-\mu^2}\frac{1}{1\mp \mathrm{ad}_X}+\frac{1}{1-\mu^2}\frac{1}{1\pm \mathrm{ad}_X}\right)\partial_\pm X \mp\frac{2\mu}{1-\mu^2}\frac{1}{1-\mathrm{ad}_X^2}\partial_\mp X \sqrt{\frac{T^{(0)}_{\pm\pm}}{T^{(0)}_{\mp\mp}}}\,.
 \end{equation}
 
\paragraph{$\mu$-frame.} Now consider the $\mu$-frame. Let us write the Lagrangian \eqref{extended_mu_L} schematically as
\begin{equation}\label{L_f_g}
    \cL=f\sqrt{T^{(0)}_{++}T^{(0)}_{--}}+g\,\mathrm{tr}(\cK_+\cO^S\cK_-)+\frac12\mathrm{tr}(\cK_+\cO^A\cK_-)+H\, ,
\end{equation}
and compute the variations with respect to $X$ of each term. Consider the following variation:
\begin{equation}
     \begin{split}
        \delta\, \tr (\cK _+ \cO_\pm\cK _-)&= -\delta\, \tr \left(\del _+X \frac 1{1\pm \ad _X}\del _-X\right)\nl 
        &=\tr \left ( \del _+\Big(\frac 1{1\pm \ad _X }\del _-X\Big)\delta X\right )+\tr \left (\del _- \Big(\frac 1 {1\mp \ad _X}\del _+ X\Big)\delta X \right ) \\
        &\pm \tr \left (\del _+ X\frac 1 {1\pm \ad _X}\delta \ad _X \frac 1 {1\pm \ad _X}\del _- X\right ). 
     \end{split}
\end{equation} 
One can then simplify the $\delta \ad_X$ term as
\begin{equation}
     \begin{split}
         \!\! \pm \tr \left (\!\!f\del _+ X\frac 1 {1\pm \ad _X}\delta \ad _X \frac 1 {1\pm \ad _X}\del _- X\!\!\right ) &\!=\! \pm f \tr \left (\!\frac 1{1\mp \ad _X}\del _+ X\, \Big[\delta X , \frac 1{1\pm \ad _X}\del_-X\Big] \!\right )\nl
         &\!=\! \mp f \tr \left (\!\Big[\frac 1{1\mp \ad _X}\del _+ X,\, \frac 1{1\pm \ad _X}\del_-X\Big]\, \delta X \!\right ) ,
     \end{split}
\end{equation}
such that finally the contribution to the EOM from this term is 
\begin{equation}
    =\del _+\left(f\frac 1{1\pm \ad _X }\del _-X\right)+\del _- \left(f\frac 1 {1\mp \ad _X}\del _+ X\right)\mp f \left [\frac 1{1\mp \ad _X}\del _+ X,\, \frac 1{1\pm \ad _X}\del_-X\right] .
\end{equation}
Now, let us look at the variations of the terms $T^{(0)}_{\pm\pm}$. Note that by using the useful property $\cO_+\cO_-=\frac12(\cO_++\cO_-)$
one only needs to compute
\begin{equation}
  \frac12 \delta (\tr (\cK _\pm\cO _-\cK _\pm))+\frac12 \delta (\tr (\cK _\pm\cO _+\cK _\pm))=\delta(\tr (\cK _\pm\cO_\mp\cO _\pm\cK _\pm)) \,\, ,
\end{equation}
and the variations of these two terms can be written as:
\bea
        \delta ( \, \tr (\cK _\pm\cO _\pm\cK _\pm)) &=&
        - \del_\pm\Big(\frac 1{1\pm\ad _X}\del _\pm X\Big)- \del_\pm\Big(\frac 1{1\mp\ad _X}\del _\pm X\Big)
        \non\\
        &&
        \pm\left[\frac 1{1\mp\ad _X}\del _\pm X,\frac 1{1\pm\ad _X}\del _\pm X\right] \,\, .
\eea
Collecting all the terms one then arrives at
\begin{equation}\label{eq:tdualeom}
    \begin{split}
        0=&-\frac12\del_+ \left (f\sqrt \frac {T^{(0)}_{--}}{T^{(0)}_{++}}\cO_+\cO_-\del _+ X+\frac12 \Big(g+\frac12\Big)\cO_-\del _- X+\frac12 \Big(g-\frac12\Big)\cO_+\del_- X\right )\nl
        &-\frac12\del_- \left (f\sqrt \frac {T^{(0)}_{--}}{T^{(0)}_{++}}\cO_+\cO_-\del _- X+\frac12 \Big(g+\frac12\Big)\cO_+\del _+ X+\frac12 \Big(g-\frac12\Big)\cO_-\del_+ X\right )\nl
        &-\frac12 \Big(g+\frac12\Big)\left [ \cO_+\del _+ X, \, \cO_-\del_-X\right ]+ \frac12 \Big(g-\frac12\Big)\left [ \cO_-\del _+ X, \, \cO_+\del_-X\right ]\nl
        &-\frac12 f\sqrt \frac {T^{(0)}_{--}}{T^{(0)}_{++}}\left[\cO_+\del _+ X,\cO_-\del _+ X\right]-\frac12 f\sqrt \frac {T^{(0)}_{++}}{T^{(0)}_{--}}\left[\cO_+\del _-X,\cO_-\del _- X\right]\,.
    \end{split}
\end{equation}
First of all notice that once we inspect the expressions for the functions $f$ and $g$ comparing \eqref{extended_mu_L} and \eqref{L_f_g}, the derivative terms match the current in \eqref{frakJ_Tdual_nu}. We have defined for short the expressions $f=\frac{\mu}{(1+\rho)^2-\mu^2}$ and $g=\frac{1}{2}\frac{-(1+\mu^2-\rho^2)}{(1+\rho)^2-\mu^2}$. Specifically, if we set $\rho=0$, the derivative terms in \eqref{eq:tdualeom} are
\begin{equation}
    \partial_+\Bigl(-\frac12 \tilde{\frak{J}}_-\Bigl)-\partial_-\Bigl(-\frac12 \tilde{\frak{J}}_+\Bigl)+\cdots
\end{equation}
To show that the EOM still coincide with the flatness of this current one needs to check that $[\tilde{\frak{J}}_+,\tilde{\frak{J}}_-]$ is twice the commutators in \eqref{eq:tdualeom}. This is exactly what is obtained by carefully computing each term of the currents' commutator, exploiting the identity $\cO_+\cO_-=\frac12(\cO_++\cO_-)$. One can hence rewrite \eqref{eq:tdualeom} as
\begin{equation}
    \partial_+\Bigl(-\frac12 \tilde{\frak{J}}_-\Bigl)-\partial_-\Bigl(-\frac12 \tilde{\frak{J}}_+\Bigl)+\frac12[\tilde{\frak{J}}_-,\tilde{\frak{J}}_+]=0\,,
\end{equation}
which indeed is twice the flatness of the current $\tilde{\frak{J}}$. Notice also that, as expected, this flatness is broken if one tries to naively keep the dependence on $\rho$. For example, in such case one would also have to include terms of the form
\begin{equation} [\tilde{\frak{J}}_+,\tilde{\frak{J}}_-]\propto\left [ \cO_\pm\del _+X, \, \cO_\pm\del_- X\right ],
\end{equation}
which do not appear in \eqref{eq:tdualeom}. These terms disappear when $\rho=0$.

\subsection{AF (bi-)Yang-Baxter deformed PCM}
\paragraph{$\nu$-frame.} For the AF (bi-)YB sigma model,  the equations of motion of the physical fields have been calculated in detail in \cite{Bielli:2024fnp}. They are written in terms of the currents
\begin{equation}
    \frak{J}_\pm=-\cO_\mp (j_\pm+2v_\pm)\,,
\end{equation}
as the following equation
\begin{equation}\label{YB_eom}
    \partial_+\frak{J}_-+\partial_-\frak{J}++\zeta\left(\left[\frak{J}_+,\tilde{\cR}\frak{J}_-\right]+\left[\tilde{\cR}\frak{J}_+,\frak{J}_-\right]\right)=0\,.
\end{equation}
Notice that in the limit $\zeta\to0$ these go back to being conservation equations for the current $\frak{J}$. Using the equations of motions for the auxiliary fields \eqref{extended_v(j,mu)} (with $\cK_\pm=j_\pm$ this time) one can again rewrite this equation in terms of
\begin{equation}
    \frak{J}_\pm=\frac{\mu^2}{1-\mu^2}\cO_\mp
j_\pm+\frac1{1-\mu^2}\cO_\pm
j_\pm-\frac{2\mu}{1-\mu^2}\left(\frac{T^{(0)}_{\pm\pm}}{T^{(0)}_{\mp\mp}}\right)^{\frac12}\cO_-\cO_+j_\mp\,.
\end{equation}
It is possible, although a bit laborious, to show that the same EOM are found by varying the Lagrangian \eqref{extended_mu_L}.

\paragraph{$\mu$-frame.}  To begin, let us note that the main difference with the T-dual model lies in the variation of the operators $\cO_\pm$. These contain a field dependent part $\cR_g=\mathrm{Ad}_{g^{-1}}\circ \cR \circ \mathrm{Ad}_g$ where $g=g(\tau,\sigma)$ is the group element defining the model. Using  the infinitesimal variation $\delta g=g \epsilon$, the variation of these objects can be computed starting from
\begin{equation}
    \delta(\mathrm{Ad}_g)X=\delta g X g^{-1}+g X \delta  g^{-1}=g\left[\epsilon,X\right]g^{-1}=\mathrm{Ad}_g\left[\epsilon,X\right] \qquad \forall \, X\in \frak{g}\,,
\end{equation}
which implies that:
\begin{equation}
    (\delta\cR_g)X= \left[\cR_g X,\epsilon\right]-\cR_g\left[X,\epsilon\right]\,.
\end{equation}
Taking into account also the variations of the inverse operator one finally obtains
\begin{equation}
    (\delta \cO_\pm) X=\mp\eta \cO_\pm\left( \left[\cR_g \cO_\pm X,\epsilon\right]-\cR_g\left[\cO_\pm X,\epsilon\right]\right)\,.
\end{equation}
At this point one has all the necessary ingredients to start computing the variations of the action \eqref{extended_mu_L}.
The two pieces that one needs to vary are:
\begin{equation}
    \sqrt{T^{(0)}_{++}T^{(0)}_{--}}=\sqrt{\mathrm{tr}((\cO_+j_+)^2)\mathrm{tr}((\cO_-j_-)^2)}\,\,,\quad \mathrm{tr}(j_+\cO^{S(A)}j_-)=\frac12\mathrm{tr}(j_+(\cO_+\pm\cO_-)j_-)\,.
\end{equation}
It can be checked that the variations produce the following equations:
\begin{align}
    \delta T^{(0)}_{\pm\pm}&=0=2\Bigl(-\partial_\pm(\cO_\mp\cO_\pm j_\pm) + [\cO_\mp\cO_\pm j_\pm,j_\pm]\nonumber\\&\mp\eta\left([\cO_\pm\cO_\mp j_\pm,\cR_g \cO_\pm j_\pm]+[\cR_g\cO_\pm\cO_\mp j_\pm, \cO_\pm j_\pm]\right)\Bigl)\,,
\end{align}
and:
\begin{align}
    \delta \,\frac12 \mathrm{tr}(j_+(\cO_+\pm\cO_-)j_-)\!=\!&+\frac12\Bigl(\!-\partial_+((\cO_+\pm \cO_-)j_-)\!-\!\partial_-((\cO_-\pm \cO_+) j_+) \!+\! [(\cO_+\pm \cO_-) j_-,j_+]\nonumber\\&+[(\cO_-\pm \cO_+) j_+,j_-]+\eta\Bigl(-[\cO_- j_+,\cR_g\cO_+j_-]+[\cO_+j_-,\cR_g\cO_-j_+]\nonumber\\
    &\pm [\cO_+j_+,\cR_g\cO_-j_-]\mp [\cO_-j_-,\cR_g\cO_+j_+]\Bigl)\Bigl)\,.
\end{align}
It is then easy to combining the above expressions to directly vary the action \eqref{extended_mu_L}:
\begin{align}\label{eom_YB_rough}
    \delta\cL=&-\frac12\partial_-\Bigg(\Big(g-\frac12\Big)\cO_+j_++\Big(g+\frac12\Big)\cO_-j_++f\left(\frac{T^{(0)}_{++}}{T^{(0)}_{--}}\right)^{\frac12}\cO_+\cO_-j_-\Bigg)\\
    &-\frac12\partial_-\Bigg(\Big(g-\frac12\Big)\cO_-j_-+\Big(g+\frac12\Big)\cO_+j_-+f\left(\frac{T^{(0)}_{--}}{T^{(0)}_{++}}\right)^{\frac12}\cO_+\cO_-j_+\Bigg)\nonumber\\
    &+f\Bigg(\Big([\cO_+\cO_- j_-,j_-]+\eta[\cO_+\cO_- j_-,\cR_g\cO_-j_-]+\eta [\cR_g\cO_+\cO_- j_-,\cO_-j_-]\Big)\left(\frac{T^{(0)}_{++}}{T^{(0)}_{--}}\right)^{\frac12}\nonumber\\
    &+\Big([\cO_+\cO_- j_+,j_+]-\eta[\cO_+\cO_- j_+,\cR_g\cO_+j_+]-\eta[\cR_g\cO_+\cO_- j_+,\cO_+j_+]\Big)\left(\frac{T^{(0)}_{--}}{T^{(0)}_{++}}\right)^{\frac12}\Bigg)\nonumber\\
    &+\frac12\Big(g+\frac12\Big)\Bigl([\cO_+ j_-,j_+]+[\cO_- j_+,j_-]-\eta [\cO_- j_+,\cR_g\cO_+j_-]-\eta [\cR_g\cO_- j_+,\cO_+j_-]\Bigl)\nonumber\\
    &+\frac12\Big(g-\frac12\Big)\Bigl([\cO_- j_-,j_+]+[\cO_+ j_+,j_-]+\eta [\cO_+ j_+,\cR_g\cO_-j_-]+\eta [\cR_g\cO_+j_+,\cO_-j_-]\Bigl)\,.\nonumber
\end{align}
Recalling that we have defined $f=\frac{\mu}{(1+\rho)^2-\mu^2}$ and $g=\frac{1}{2}\frac{-(1+\mu^2-\rho^2)}{(1+\rho)^2-\mu^2}$, from the above result one can see that the expressions acted upon by derivatives exactly coincide with one-half of the $\frak{J}_\pm$ current found in \eqref{YB_eom}. Therefore one needs to show that the remaining commutator terms in \eqref{eom_YB_rough} can be rearranged into the form $-\frac12\zeta([\frak{J}_+,\tilde\cR\frak{J}_-]+[\frak{J}_-,\tilde\cR\frak{J}_+])$. When this is true, then the $\mu$-frame equations of motion are proportional, with an overall factor $\frac12$, to the equation \eqref{YB_eom}. It is convenient to proceed by expanding the latter two commutators involving $\mathfrak{J}$ and checking that they lead to the same commutator terms appearing in \eqref{eom_YB_rough}.
Assuming that $\frak{J}_\pm$ take the form
\begin{equation}
    \frak{J}_\pm=b\cO_\pm j_\pm+a\cO_\mp j_\mp+2f\left(\frac{T^{(0)}_{\pm\pm}}{T^{(0)}_{\mp\mp}}\right)^{\frac12}\cO_+\cO_-j_\mp\,,\quad \text{with }\quad a=g+\frac12\,,\,b=g-\frac12\,,
\end{equation}
one can indeed show that the commutators
\begin{equation}\label{zeta_commutators}
    -\zeta([\frak{J}_+,\tilde\cR\frak{J}_-]+[\frak{J}_-,\tilde\cR\frak{J}_+])\end{equation}
are precisely equal to all the commutators in \eqref{eom_YB_rough} provided that $\rho=0$, as it was the case for the T-dual model, which allows to simplify the calculation using the relations
\begin{equation}
\begin{aligned}
a-b=1\,,\quad\qquad\,\, ab-f^2&=0\,,\qquad\qquad a^2-ab=a\,,
\\
ab-b^2=b 
    \qquad\qquad a^2-f^2&=a\,,\qquad\qquad  b^2-f^2=-b\,.
\end{aligned}
\end{equation}
When $\rho=0$ these reduce to $a=\frac{-\mu^2}{1-\mu^2}$ and $b=-\frac{1}{1-\mu^2}$. It is finally useful to note that the following identities can be used along the computation
\begin{equation}
    \zeta \tilde\cR\cO_\pm=(M-\eta\cR_g)\cO_\pm\,,\quad M\cO_\pm=\pm(\mathds1-\cO_\pm)\,,\quad \cO_+\cO_-=\frac12(\cO_++\cO_-)\,.
\end{equation}
Using all the above ingredients it is possible to reduce the $14$ commutators in \eqref{eom_YB_rough} to the two commutators \eqref{zeta_commutators} up to a factor of $\frac12$. 

\bibliographystyle{utphys}
\bibliography{master}
\end{document}